\documentclass[preprint]{aastex}

\usepackage{CJK}
\usepackage{subfigure}
\usepackage{float}
\usepackage{color}
\usepackage{hyperref}
\usepackage{amsmath}

\slugcomment{September 8, 2011}

\def\vycma   {VY CMa}
\def\Jthron  {J0731$-$2341}
\def\Jtwozo  {J0720$-$2628} % observed only at the first epoch
\def\Jtwofo  {J0724$-$2515} % observed only at the first epoch
\def\Jtwofi  {J0725$-$2640}

\def\hho     {H$_2$O}

\def\vlacal  {J0730$-$1141}

\def\VLSR    {$V_{\rm LSR}$}
\def\VLSRS   {$V_{\rm LSR}*$}
\def\kms     {km~s$^{-1}$}
\def\masy    {mas~yr$^{-1}$}
\def\mjybeam {mJy~beam$^{-1}$}
\def\jybeam  {Jy~beam$^{-1}$}

\def\msun    {$M_{\odot}$}

\def\dx      {$\Delta x$}
\def\dy      {$\Delta y$}
\def\ra      {$\alpha_{\rm J2000}$}
\def\dec     {$\delta_{\rm J2000}$}

\def\h       {\ifmmode{^{\rm h}}\else{$^{\rm h}$}\fi}
\def\m       {\ifmmode{^{\rm m}}\else{$^{\rm m}$}\fi}
\def\s       {\ifmmode{^{\rm s}}\else{$^{\rm s}$}\fi}
\def\deg     {\ifmmode{^{\circ}}\else{$^{\circ}$}\fi}
\def\decdeg  {\ifmmode{{\rlap.}^{\circ}} \else ${\rlap.}^{\circ}$\fi}
\def\decs    {\ifmmode{{\rlap.}^{\rm s}} \else ${\rlap.}^{\rm s}$\fi}
\def\decas   {\ifmmode{{\rlap.}{''}}\else{${\rlap.}{''}$}\fi}

\def\Vsbar {\ifmmode {\overline{V_s}}\else {$\overline{V_s}$}\fi}
\def\Usbar {\ifmmode {\overline{U_s}}\else {$\overline{U_s}$}\fi}
\def\Wsbar {\ifmmode {\overline{W_s}}\else {$\overline{W_s}$}\fi}

\def\mux    {\ifmmode {\mu_x}\else {$\mu_x$}\fi}
\def\muy    {\ifmmode {\mu_y}\else {$\mu_y$}\fi}
\def\mura   {\ifmmode {\mu_{\alpha}}\else {$\mu_{\alpha}$}\fi}
\def\mude   {\ifmmode {\mu_{\delta}}\else {$\mu_{\delta}$}\fi}

\begin{document}

\begin{CJK*}{UTF8}{gbsn}
\title{Distance and Kinematics of the Red Hypergiant \vycma: VLBA and VLA Astrometry}

\author{B. Zhang (张波)\altaffilmark{1,5},  M. J. Reid\altaffilmark{2}, K. M.
Menten\altaffilmark{3}, X. W. Zheng (郑兴武)\altaffilmark{4}} 

\altaffiltext{1}{Shanghai Astronomical Observatory, Chinese Academy of
Sciences, Shanghai 200030, China}

\altaffiltext{2}{Harvard-Smithsonian Center for Astrophysics, 60
 Garden Street, Cambridge, MA 02138, USA}

\altaffiltext{3}{Max-Plank-Institut f\"ur Radioastronomie, Auf dem H\"ugel
69, 53121 Bonn, Germany}

\altaffiltext{4}{Department of Astronomy, Nanjing University, Nanjing
    210093, China}

\altaffiltext{5}{Now at: Max-Plank-Institut f\"ur Radioastronomie, Auf dem H\"ugel
69, 53121 Bonn, Germany; bzhang@mpifr.de}

\begin{abstract}

We report astrometric results of phase-referencing VLBI observations of
43 GHz SiO maser emission toward the red hypergiant VY Canis Majoris
(\vycma) using the Very Long Baseline Array (VLBA).  We measured a
trigonometric parallax of 0.83 $\pm$ 0.08 mas, corresponding to a
distance of 1.20$^{+0.13}_{-0.10}$ kpc.  Compared to previous studies,
the spatial distribution of SiO masers has changed dramatically, while
its total extent remains similar.  The internal motions of the maser
spots are up to 1.4 \masy, corresponding to 8 \kms, and show a tendency
for expansion.  After modeling the expansion of maser spots, we derived
an absolute proper motion for the central star of \mux\ = $-2.8 \pm 0.2$
and \muy\ = $2.6 \pm 0.2$ \masy\ eastward and northward, respectively.
Based on the maser distribution from the VLBA observations, and the
relative position between the radio photosphere and the SiO maser
emission at 43 GHz from the complementary Very Large Array (VLA)
observations, we estimate the absolute position of \vycma\ at mean epoch
2006.53 to be \ra\ = 07\h 22\m 58\decs3259 $\pm$ 0\decs0007, \dec\ =
$-$25\arcdeg 46\arcmin03\decas063 $\pm$ 0\decas010.  The position and
proper motion of \vycma\ from the VLBA observations differ significantly
with values measured by the Hipparcos satellite.  These discrepancies
are most likely associated with inhomogeneities and dust scattering the
optical light in the circumstellar envelope.  The absolute proper motion
measured with VLBA suggests that \vycma\ may be drifting out of the
giant molecular cloud to the east of it.

\end{abstract}


\section{INTRODUCTION}

VY Canis Majoris (\vycma) is one of most massive and luminous red
hypergiant stars in our Galaxy.  At its ``traditional'' estimated
distance of 1.5 kpc, based on the angular proximity to the cluster NGC
2362~\citep{lada78}, \vycma\ would be near the empirical upper
luminosity boundary in the Hertzsprung-Russel (H-R)
diagram~\citep{schuster06}, associated with high mass-loss and ejection
phenomena.  Because the red hypergiant phase represents a very
short-lived evolutionary stage, with a timescale of only $\sim10^5$
years, the physical properties and evolutional state of this object have
been investigated extensively from optical and infrared to radio
wavelengths.

\vycma\ displays OH, \hho\ and SiO maser emission in its circumstellar
envelope (CSE).  Very Long Baseline Interferometry (VLBI) observations
of SiO masers from oxygen-rich asymptotic giant branch stars (e.g., Mira
variables) have shown that the masers are situated only a few radii from
the stellar surface, giving us insight into the properties of the inner
CSE, which is important for the study of the evolution of late type
stars~\citep{boboltz05}.  Investigations of the distributions of the
SiO masers show that the masers typically form ring-like
structures~\citep{diamond94,greenhill95,cotton04}, possibly with a
rotational component
~\citep{boboltz00a,boboltz00b,hollis01,sanchez02}.

VLBI observations of SiO maser emission from \vycma\ in the $J = 1 \to
0$, $v = 1$ and $v = 2$ transitions showed them to be approximately
spatially coincident, with extents of approximately 80~mas on the sky
and the strongest spots concentrated toward the east~\citep{miyoshi94,
miyoshi03}.  VLBI maps of the $v = 1$, $J = 2 \to 1$ maser emission
extend over about 100 $\times$ 80 mas in right ascension and
declination, respectively, estimated at about 2$-$4 stellar
radii~\citep{shibata04}. 

Knowledge of distance is very important to determine the physical
properties of a star.  Inferred stellar radii depend directly on
distance and luminosities scale as the square of distance; moreover,
distances indirectly affect modeled stellar properties, such as
effective temperature and surface gravity.  Recently, \citet{choi08a}
measured a trigonometric parallax for \vycma\ of $0.88 \pm 0.08$ mas,
corresponding to a distance of $1.14^{+0.11}_{-0.09}$ kpc, based on
observations of \hho\ masers with the Japanese VLBI Exploration of Radio
Astrometry (VERA) array.  At this distance, \vycma\ more comfortably
falls below the theoretical maximum luminosity on the HR diagram.  Our
parallax measurements presented here are made with a different
telescope, the Very Long Baseline Array (VLBA), observing a different
maser species (SiO) and thus give an independent measurement of the
distance and proper motion of \vycma.

While stellar positions for red giant stars have been measured with the
Hipparcos satellite, these are generally far less accurate than for
dwarf stars, since red giant stars are large, variable, often surrounded
by copious dust and far away.  Another way to locate a red giant star is
to observe circumstellar masers with radio
interferometry~\citep{baudry84}.  Because masers usually surround the
red giant star, the radio position of the central star can be estimated
from the observed distribution and kinematics of the SiO masers, which
can be linked to the positions of extragalactic quasars by VLBI
phase-referencing~\citep{colomer93}.  More directly, however, radio
continuum emission from the stellar photosphere can be imaged by, for
example, the Very Large Array (VLA) using \hho\ or SiO masers as a
phase-reference~\citep{reid90, reid97, reid07}, which provides a
direct and accurate measurement of the maser distribution relative to
the star.

In this paper, we present the results from our multi-epoch radio
interferometer observations of the SiO maser emission and radio
continuum toward \vycma.  In \S\ref{sec:obs}, we describe the VLBA
phase-referencing observations of SiO masers and the VLA observations of
weak radio photospheric continuum and maser emission at 43 GHz made on
the day after the second epoch of our VLBA observations.  In
\S\ref{sec:int_mot}, we compare the spatial distribution of SiO masers
at different epochs and estimate their internal motions. In
\S\ref{sec:para}, we use the time variation of the positions of maser
spots relative to a background source to determine a trigonometric
parallax and absolute proper motion.  In \S\ref{sec:abs_pos}, we
describe the procedures to determine the absolute position of the
central star based on the quasi-ring-like structure of maser
distributions from the VLBA observations and the VLA observations of
radio photospheric emission using strong maser emission as
phase-reference.  In \S\ref{sec:kin_mod}, after modeling the kinematics
of masers, we derive an absolute proper motion of the star.  In
\S\ref{sec:disc}, we compare the positions and proper motions with
values measured by the Hipparcos satellite and VERA, and discuss the
reasons for the difference and the origin of \vycma.

\section{OBSERVATIONS} \label{sec:obs} 
\subsection{VLBA Observations}

Our VLBI observations were conducted with the VLBA operated by the
National Radio Astronomy Observatory (NRAO)\footnote{The National Radio
Astronomy Observatory is a facility of the National Science Foundation
operated under cooperative agreement by Associated Universities, Inc.}
under program BR106. We observed the $v = 1$, $J = 1 \to 0$ SiO maser
transition at a rest frequency of 43.12208 GHz toward \vycma\ with
8-hour tracks on 2005 October 20, 2006 April 16 and September 29, and
2007 April 14.  This time sampling provides nearly maximum sensitivity
for parallax detection and ensures that we can separate the secular
proper motion from the sinusoidal parallax effect.  We scheduled the
observations so as to maximize the right ascension (and not declination)
parallax offsets for two reasons.  First, for \vycma\ the amplitude of
the parallax signature in right ascension is greater than in
declination.  Second, since \vycma\ is observed at low elevation angles,
the uncertainty of declination measurements would be expected to be
considerably larger than for right ascension \citep{Honma08}.

We observed several extragalactic radio sources as potential background
references for parallax solutions.  The observing sequence was \vycma,
\Jthron, \vycma, \Jtwozo, \vycma, \Jtwofo, \vycma, \Jtwofi\ for the
first epoch.  From the first epoch observations, we found \Jtwozo\ and
\Jtwofo\ too weak to be detected, and we dropped these sources from the
remaining observations.  We switched between the maser target and
background sources every 40~s, typically achieving 30~s of on-source
data.  We used an SiO maser spot as the phase-reference source because
it is considerable stronger than the background source and could be
detected on individual baselines in the available on-source time.
Table~\ref{tab:src_pos} lists the positions, intensities, source
separations, LSR velocity of the reference maser feature and synthesized
beam sizes. 

\begin{table}[H]
  \footnotesize
  \caption[]{Positions and Brightnesses \label{tab:src_pos}}
  \begin{center}
  \begin{tabular}{cccccccl}
\hline \hline
Source      &  R.A. (J2000)  & Dec. (J2000)   & $S_p$    &  $\theta_{sep}$   & P.A. & \VLSR & Beam \\
            & (h~~~m~~~s)  & (\degr~~~\arcmin~~~\arcsec) & (Jy/beam) &  (\degr) & (\degr)  & (\kms) & (mas~~mas~~\degr) \\
\hline
\vycma.......&  07 22 58.3283 & $-$25 46 03.075   &  18 $-$ 27     &  ...  & ...      & 33.3 & 0.5 $\times$ 0.2 @ $-$12\\
\Jtwofi...   &  07 25 24.4130 & $-$26 40 32.680   &   0.03         &  1.1  & $+$ 34   & ...  & 2.7 $\times$ 0.6 @ $-$11\\
\Jthron...   &  07 31 06.6680 & $-$23 41 47.869   &   0.06         &  2.8  & $+$136   & ...  & 0.8 $\times$ 0.3 @ $+$6\\
\hline
  \end{tabular}
  \end{center}
\small{
Note. --- The fourth and seventh columns give the peak brightnesses ($S_p$) and
\VLSR\ of reference feature.  The fifth and sixth columns give the separations
($\theta_{sep})$ and position angles (P.A.) east of north between maser and
background sources.  The last column gives the FWHM size and P.A. of the
Gaussian restoring beam.  Calibrator \Jtwofi~is from the VLA program AR569, and
its position is corrected using information from the VLBA Calibrator Survey
(VCS) by \citet{kovalev07}, the information for \Jthron~is from VCS
by~\citet{petrov05}.  
}
\end{table}

We placed observations of two strong sources (J0730$-$1141 and
J0530$+$1131) near the beginning, middle, and end of the observations in
order to monitor delay and electronic phase differences among the
intermediate-frequency bands. The rapid-switching observations employed
two adjacent bands of 8~MHz bandwidth and recorded both right and left
circularly polarized signals. The two (dual-polarized) bands were
centered at Local Standard of Rest velocities (\VLSR) of $-33.6$ and
$22.0$ \kms\ for \vycma. The SiO masers were contained in the second
band.

In order to do atmospheric delay calibration, we placed ``geodetic"
blocks before and after our phase-reference observations
\citep{reid09a}.  These data were taken in left circular polarization
with eight 8~MHz bands that spanned 480~MHz of bandwidth between 42.9
and 43.4~GHz; the bands were spaced in a ``minimum redundancy
configuration" to uniformly sample, as best as possible, all frequency
differences. 

The data were correlated in two passes with the VLBA correlator in
Socorro, NM.  One pass generated 16 spectral channels for all the data
and a second pass generated 256 spectral channels, but only for the
single (dual-polarized) frequency band containing the maser signals,
giving a velocity resolution of 0.21 \kms.  The data calibration was
performed with the NRAO AIPS package as described by \citet{reid09a}.

\subsection{VLA Observations}

Observation of the radio photosphere and maser emission toward \vycma\
at 43 GHz were made with VLA in its largest (A) configuration under
program AR595 on 2006 April 17.  We used a dual intermediate frequency
band setup with a narrow (6.25 MHz) band covering an LSR velocity range
of $-5$ to $40$ \kms\ for the $v = 1, J = 1 \to 0$ SiO maser line (rest
frequency of 43122.08 MHz) and a broad (50 MHz) IF band centered
$\approx50$ MHz above the masers on a line-free portion of the spectrum.
We also observed the SiO maser emission at high spectral resolution with
several scans in spectral-line mode interspersed among the dual-band
continuum observations.  These covered the SiO line emission with a
bandwidth of 6.25 MHz centered at 22 \kms\ using 128 spectral channels,
which provided a channel spacing of 0.34 \kms. 

A typical observing unit for our continuum observations consisted of a
$\sim50$ minute \vycma\ scan, followed by a $\sim$ 1.5 minute scan of,
alternately, the quasar J0730--1141 and J0648--3044.  For a
spectral-line observation, we observed a calibrator \vlacal\ for about
1.5 minutes and \vycma\ for 6.5 minutes.  Absolute flux density
calibration was established by an observation of 3C~286 (J1331+3030),
assuming a flux density of 1.46 Jy at 43 GHz.  Data calibration
procedures are described in detail in \citet{reid97, reid07}.

\section{SPATIAL DISTRIBUTION AND INTERNAL MOTION OF MASER SPOTS}
\label{sec:int_mot}

\begin{figure}[H]
  \centering
  \includegraphics[angle=0,scale=0.80]{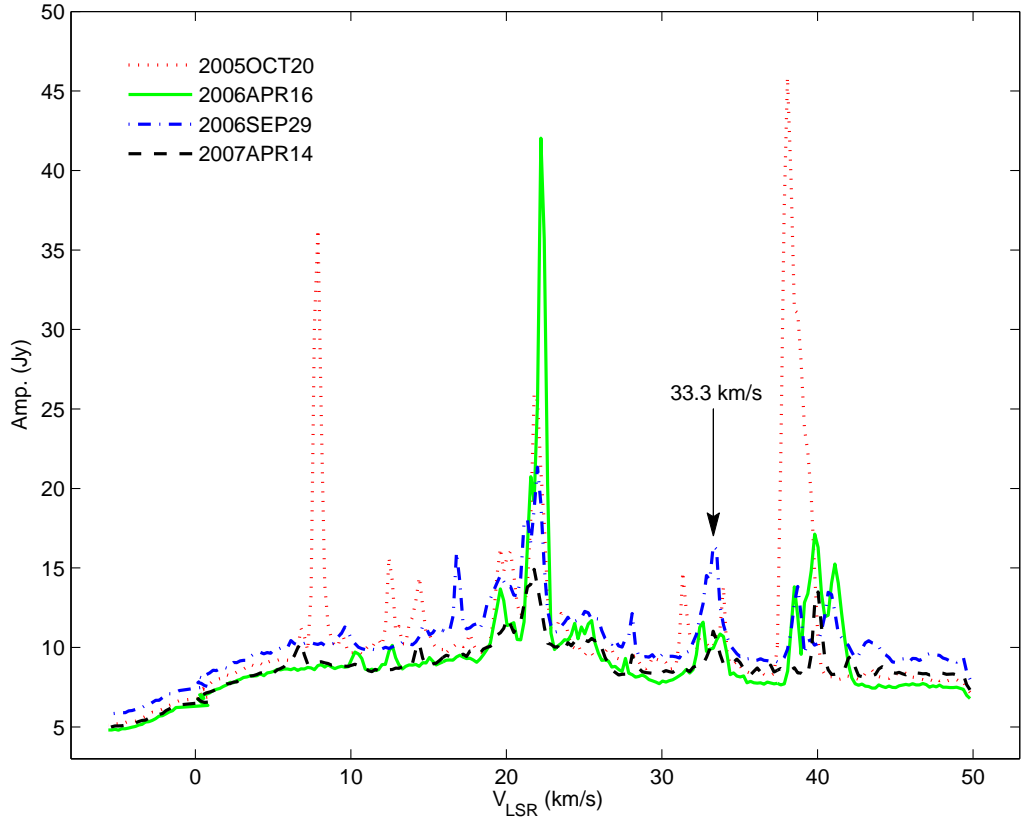}
  \caption{
Interferometer (scalar averaged cross-power amplitude over the full
duration of the observation) spectra of the SiO masers toward \vycma\
obtained with the long VLBA baseline from Pie Town (New Mexico) to Mauna
Kea (Hawaii) at four epochs.  The {\it arrow} points to the maser
feature at \VLSR\ of 33.3~\kms\ which served as the phase-reference. 
\newline (A color version of this figure is available in the online journal.)
} 
  \label{fig:vycma_spec} 
\end{figure}

Fig.~\ref{fig:vycma_spec} shows interferometer spectra of SiO maser
emission observed with a long VLBA baseline at all four epochs.  The SiO
maser emission spans a \VLSR\ range of about 6 to 42~\kms.  One can see
that the flux densities of some maser features varied considerably over
1.5~yr.  We selected a compact and relative stable strong maser spot at
\VLSR\ of 33.3~\kms\ to serve as the phase-reference.  The point-source
response function (dirty beam) typically had a FWHM of 0.5 by 0.2~mas at
a position angle of $-12$\deg\ east of north using uniform weighting. 

%--- Fig- spectrum 
\clearpage

%--- Fig- maser distribution (momnt0)
\clearpage
\begin{figure}[H]
  \centering
  \includegraphics[angle=-90,scale=0.85]{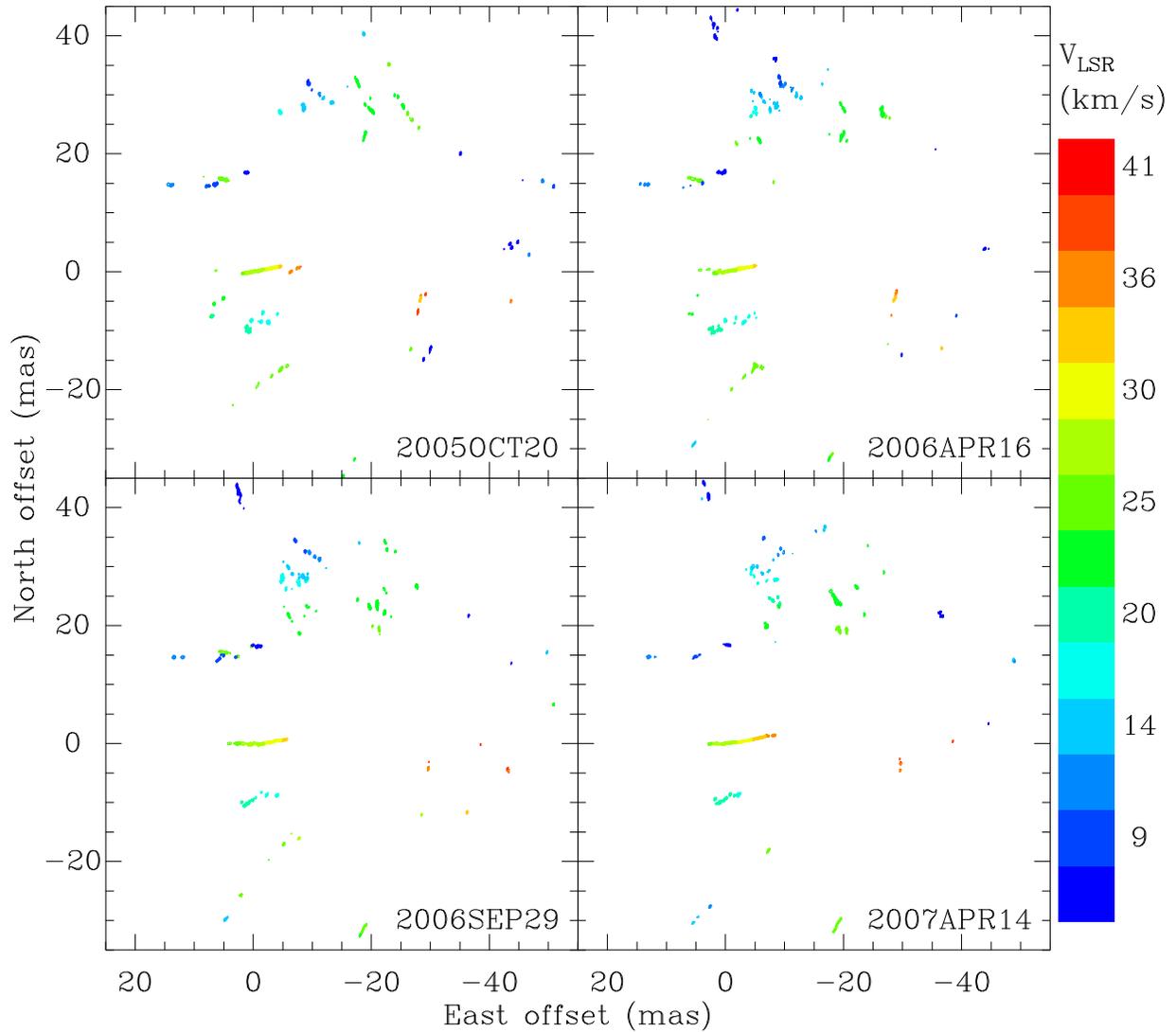}
  \caption{
Spatial distribution of SiO maser emission towards \vycma\ observed at
four epochs.  Observation dates are indicated in the lower right corner
of each panel.  The \VLSR\ velocity of the maser spots is color-coded as
indicated by the color bar on the right panel. 
\newline (A color version of this figure is available in the online journal.)
  \label{fig:vycma_maser_pos}} 
\end{figure}
\clearpage

\begin{figure}[H]
  \centering
  \includegraphics[angle=0,scale=0.80]{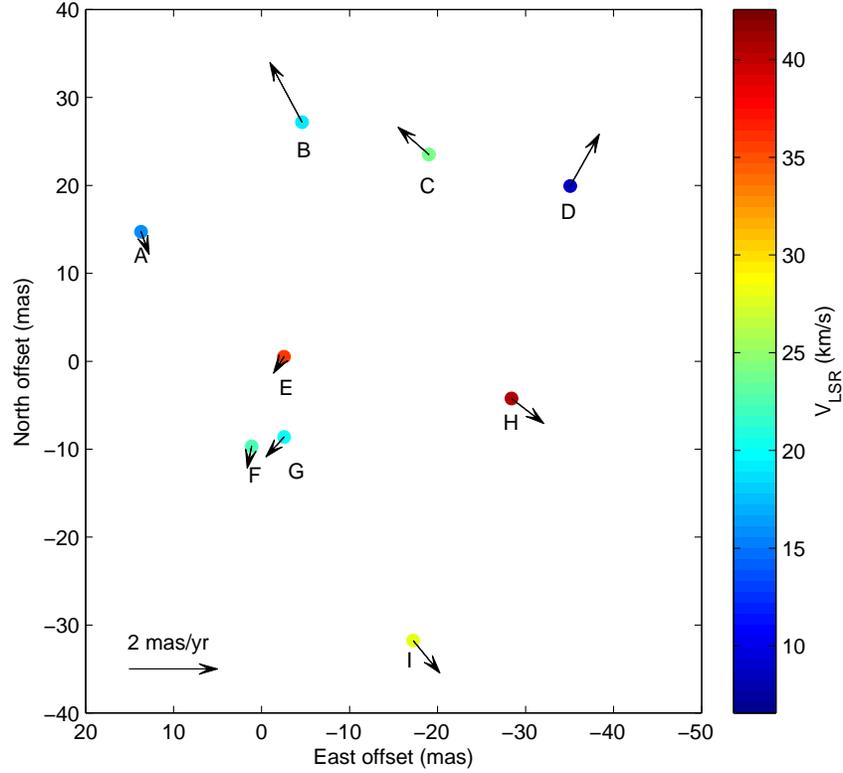}
  \caption{
Averaged positions ({\it circle and letter label}) and internal motions
({\it arrow}) of the SiO maser features towards \vycma. The average
motion of all features has been removed. The color bar denotes the
\VLSR\ range from 42 to 6 \kms\ of the maser features.  The length and
the direction of an {\it arrow} indicate the speed (given by the scale
{\it arrow} in the lower left of the figure) and the direction of
internal motion of a maser feature.  The reference maser spot is in
feature E, located at (0,0) mas, which is slightly different from the
average position of maser spots in feature E. 
\newline (A color version of this figure is available in the online journal.)
  \label{fig:maser_pos_pm}} 
\end{figure}

Figure~\ref{fig:vycma_maser_pos} shows the spatial distribution of SiO
maser emission towards \vycma\ relative to the reference maser spot at
\VLSR = 33.3~\kms.  The total extent of the SiO maser spot distribution
is about 80~mas, which is consistent with previous
studies~\citep{miyoshi94, miyoshi03}, while the spatial distribution
changes in detail.  We considered maser spots at different epochs as
being from the same feature if their positions in the same spectral
channel were coincident within 5 mas at different epochs, corresponding
to a motion of less than 20 \kms.  Selecting only the brightest maser
spot in each channel, we found nine features  including at least two
maser spots in adjacent channels that were detected at all four epochs
(see Fig.~\ref{fig:maser_pos_pm}).  

The relative motions of features were calculated by performing a
weighted least-squares fit for linear motion with respective to the
reference feature.  After removing the average relative motion of all
maser features, the remaining (internal) motion vectors, indicated with
arrows in Fig.~\ref{fig:maser_pos_pm}, exhibit an expanding structure.
These internal motions are also listed in Table~\ref{tab:avg_ipm}; the
maximal motion is measured for feature B.  It is about 1.4 \masy,
corresponding to 8 \kms\ at a distance of 1.2 kpc. 

\begin{figure}[H]
  \centering
  \epsscale{.9}
  \plottwo{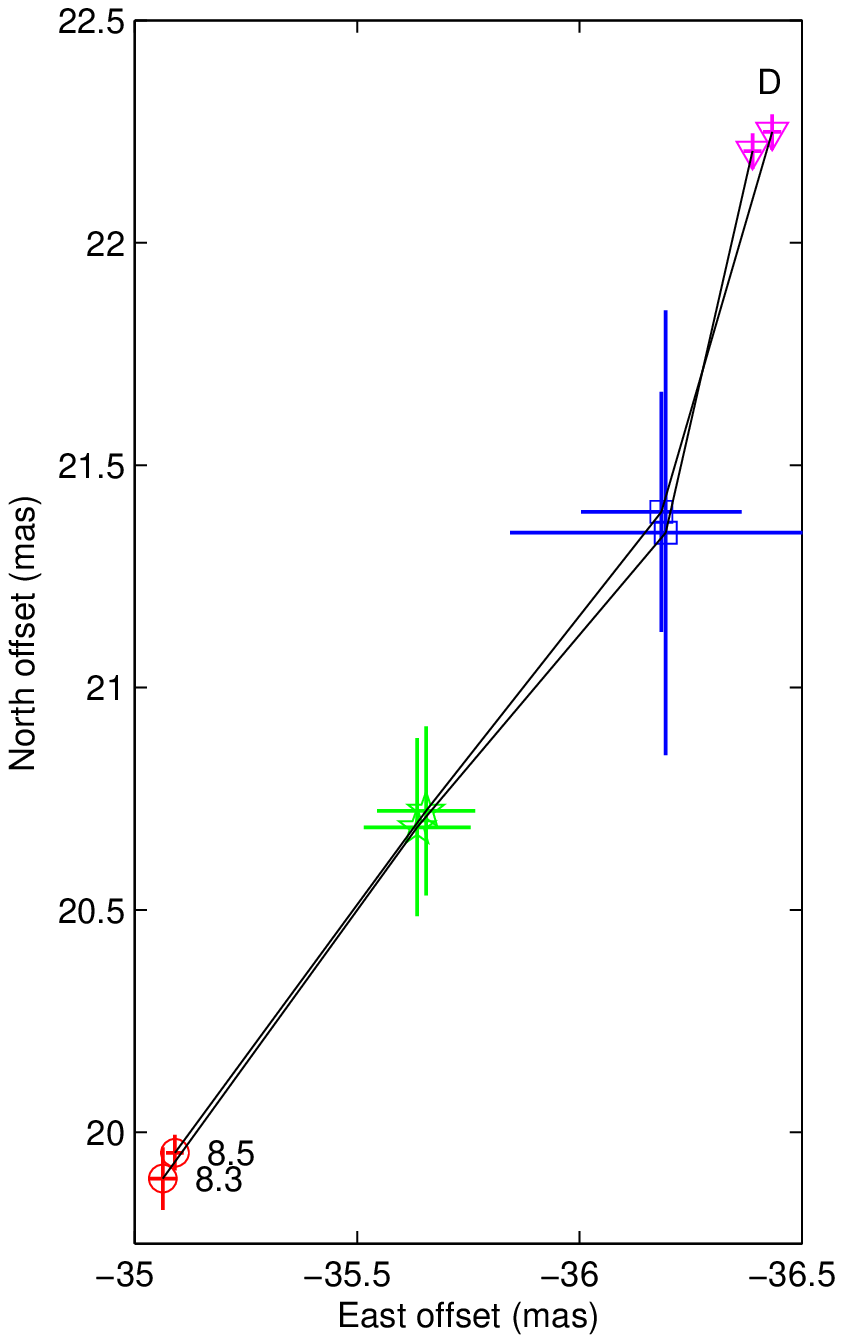}{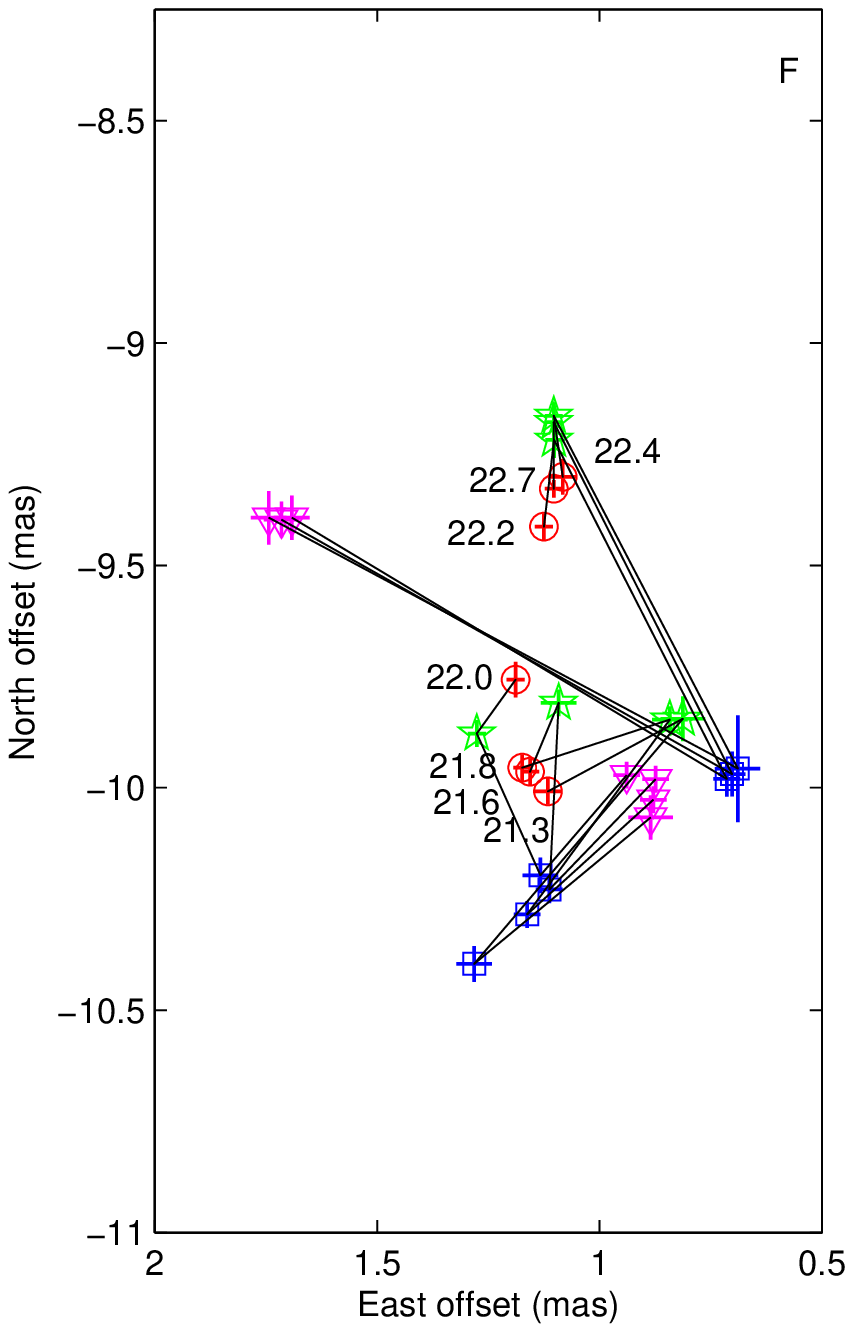}
  \caption{
Temporal changes of relative positions of the SiO maser spots in feature
D and F.  {\it Circles, pentagons, squares} and {\it triangles} denote
the first to the last epoch, respectively.  The error bars are 10 times
the formal position uncertainty.  The \VLSR\ of each maser spot is
indicated by the number close to the position at the first epoch. 
\newline (A color version of this figure is available in the online journal.)
  \label{fig:reg_im}} 
\end{figure}

Fig.~\ref{fig:reg_im} shows the time variations of the positions of some
maser spots relative to the reference spot.  The masers in feature D
exhibit approximate linear motions, while those in feature F show less
regular motions.  Indeed, in feature F, it is difficult to avoid
possible misidentification of the same maser spots from one epoch to
another, which introduces uncertainty in the kinematic parameters of
those maser spots.  Thus, for parallax measurements, only those with
linear motions were used.

%--- Fig- maser distribution and internal motion
\clearpage

\clearpage

\begin{deluxetable}{crrr}
%\tabletypesize{\scriptsize}
%\rotate
\tablecaption{Internal motions of maser features \label{tab:avg_ipm}}
\tablewidth{0pt}
\tablehead{
\colhead{Region}&\colhead{\VLSR}  &\colhead{\mux}  & \colhead{\muy} \\
\colhead{}      &\colhead{(\kms)}   &\colhead{(\masy)} & \colhead{(\masy)} 
} 
\startdata
     A & 15.0 & -0.187 $\pm$  0.029 &-0.511 $\pm$  0.003 \\ 
     B & 18.5 &  0.732 $\pm$  0.087 & 1.359 $\pm$  0.007 \\ 
     C & 23.7 &  0.705 $\pm$  0.202 & 0.621 $\pm$  0.055 \\ 
     D &  8.5 & -0.664 $\pm$  0.007 & 1.182 $\pm$  0.003 \\ 
     E & 35.5 &  0.240 $\pm$  0.013 &-0.377 $\pm$  0.011 \\ 
     F & 22.0 &  0.100 $\pm$  0.035 &-0.501 $\pm$  0.011 \\ 
     G & 19.8 &  0.418 $\pm$  0.007 &-0.455 $\pm$  0.012 \\ 
     H & 39.8 & -0.735 $\pm$  0.070 &-0.573 $\pm$  0.057 \\ 
     I & 27.6 & -0.610 $\pm$  0.114 &-0.745 $\pm$  0.143 \\ 
\enddata
\small{
Note. --- Motions $\mux = \mu_{\alpha \cos{\delta}}$ and $\muy = \mu_{\delta}$ are 
listed with the unweighted average motion removed.
}
\end{deluxetable}

\clearpage

\clearpage

\section{PARALLAX AND PROPER MOTION}
\label{sec:para}

We fitted elliptical Gaussian brightness distributions to the images of
strong maser spots and the extragalactic radio sources for all four
epochs.  The change in position of each maser spot relative to each
background radio source was modeled by the parallax sinusoid in both
coordinates (determined by a single parameter, the star's parallax) and
a linear proper motion in each coordinate.

\begin{figure}[H]
  \centering
  \includegraphics[angle=-90,scale=0.70]{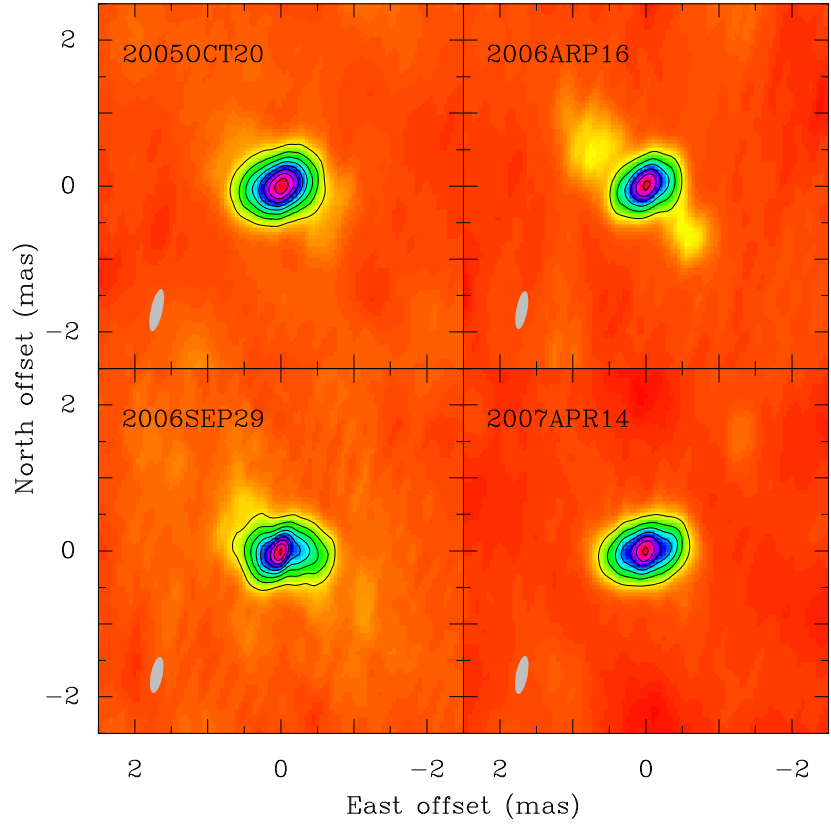}
  \caption{
Images of the reference SiO maser spot at \VLSR\ = 33.3 \kms\ in \vycma.
Observation dates are indicated in the upper left corner and the
restoring beam (gray) is indicated in the lower left corner of each
panel. Contour levels are spaced linearly at 3.0 \jybeam.
\newline (A color version of this figure is available in the online journal.)
  \label{fig:vycma_maser}}
\end{figure}

Fig.~\ref{fig:vycma_maser} shows the maser reference channel images at
all four epochs.  One can see that the emission appears dominated by a
single compact component, and there is no dramatic variation over the
1.5 year time span of our observations.  The extragalactic source
\Jthron\ was relatively far (2.8\deg) from \vycma, and it was only used
to determine the absolute position of the maser reference spot.   We
only used \Jtwofi, separated by 1.1\deg\ from \vycma, to determine the
parallax.

\begin{figure}[H]
  \centering
  \includegraphics[angle=-90,scale=0.90]{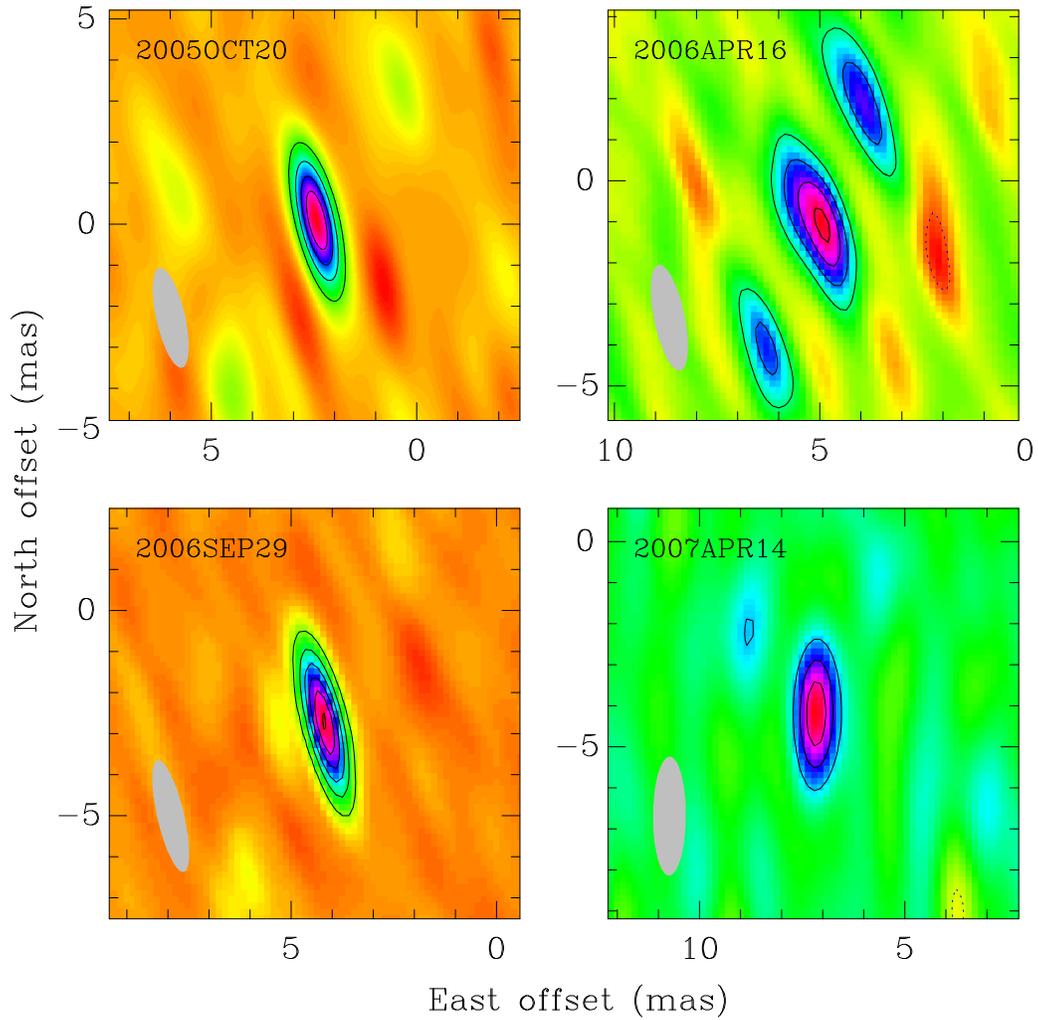}
  \caption{
Images of extragalactic radio source \Jtwofi\ used for the parallax
measurements of \vycma.  Observation dates are in the upper left corner
and the restoring beam (gray) is in the lower left corner of each panel.
Contour levels are spaced linearly at 25, 25, 10 and 5 \mjybeam\ for the
four epochs in chronological order.
\newline (A color version of this figure is available in the online journal.)
  \label{fig:vycma_qso}}
\end{figure}

Fig.~\ref{fig:vycma_qso} provides images of background radio source
\Jtwofi\ using only the inner 5 VLBA antennas at all epochs.  This
source is dominated by a single component with peak brightnesses of
0.13, 0.04, 0.15 and 0.02 \mjybeam\ at our four epochs.  The variation
of the brightnesses is probably caused in part by significant flux loss
owing to poor phase coherence in the two April observations, which were
in evening hours when coherence is often poor.

%--- Fig- image of reference maser source
\clearpage

%--- Fig-QSO
\clearpage

\begin{figure}[H]
  \centering
    \includegraphics[angle=-90,scale=0.65]{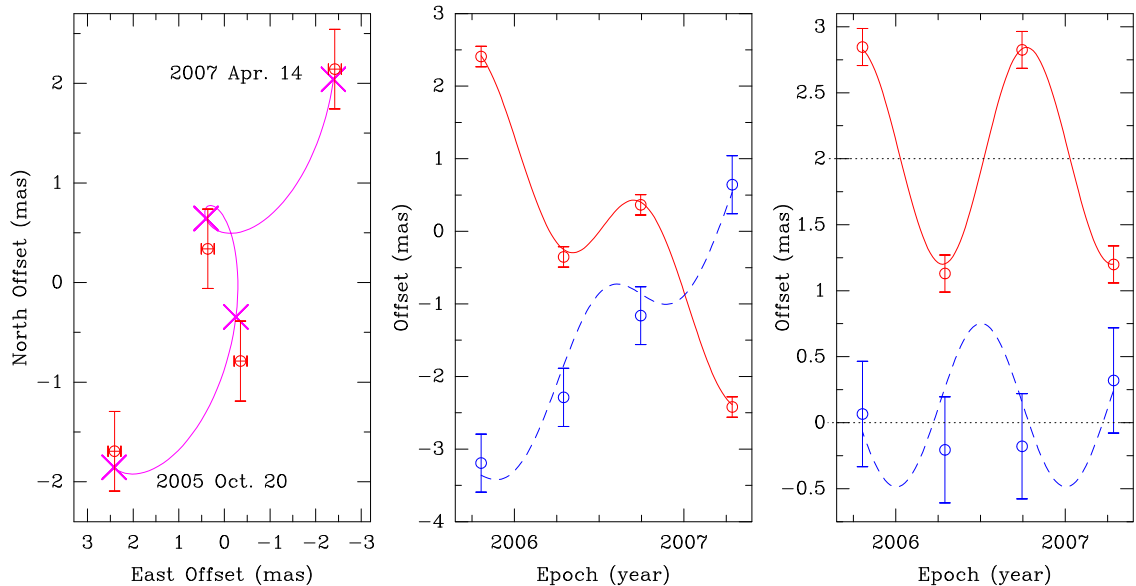}
    \caption{
Parallax and proper motion data ({\it circles}) and best-fitting models
({\it lines}) for the reference maser spot at \VLSR\ of 33.3~\kms.
Plotted are positions of the maser spot relative to the extragalactic
radio source \Jtwofi.  {\it Left Panel}: Positions on the sky for the
first and last epochs are labeled.  The expected positions from the
parallax and proper motion fit are indicated with ({\it crosses}). {\it
Middle Panel}: Eastward ({\it solid lines}) and northward ({\it dashed
lines}) offsets and best-fitting models versus time.  Data for the
eastward and northward positions are offset vertically for clarity.
{\it Right Panel}: Same as the {\it middle panel}, except the
best-fitting proper motion has been removed, displaying only the
parallax signature.
\newline (A color version of this figure is available in the online journal.)
    \label{fig:vycma_para}} 
\end{figure}

As mentioned in \S\ref{sec:int_mot}, the apparent motions of the maser
spots can be complicated by a combination of spectral blending and
changes in intensity.  Thus, for parallax fitting, one needs to find
stable, unblended spots and/or use many maser spots.  Thus, we first
fitted a parallax and proper motion to the position offsets for each
maser spot separately (with respect to the background radio
source~\Jtwofi).  In Fig.~\ref{fig:vycma_para}, we plot the position of
the reference maser spot relative to the background radio source as an
example, with superposed curves representing the model maser tracks
across the sky.  This is one of the most compact and unblended spots
that was detectable at all four epochs.

%--- Fig-  for figure of parallax curve of maser spot from reference channel 
\clearpage

\begin{figure}[H]
  \centering
    \includegraphics[angle=0,scale=0.8]{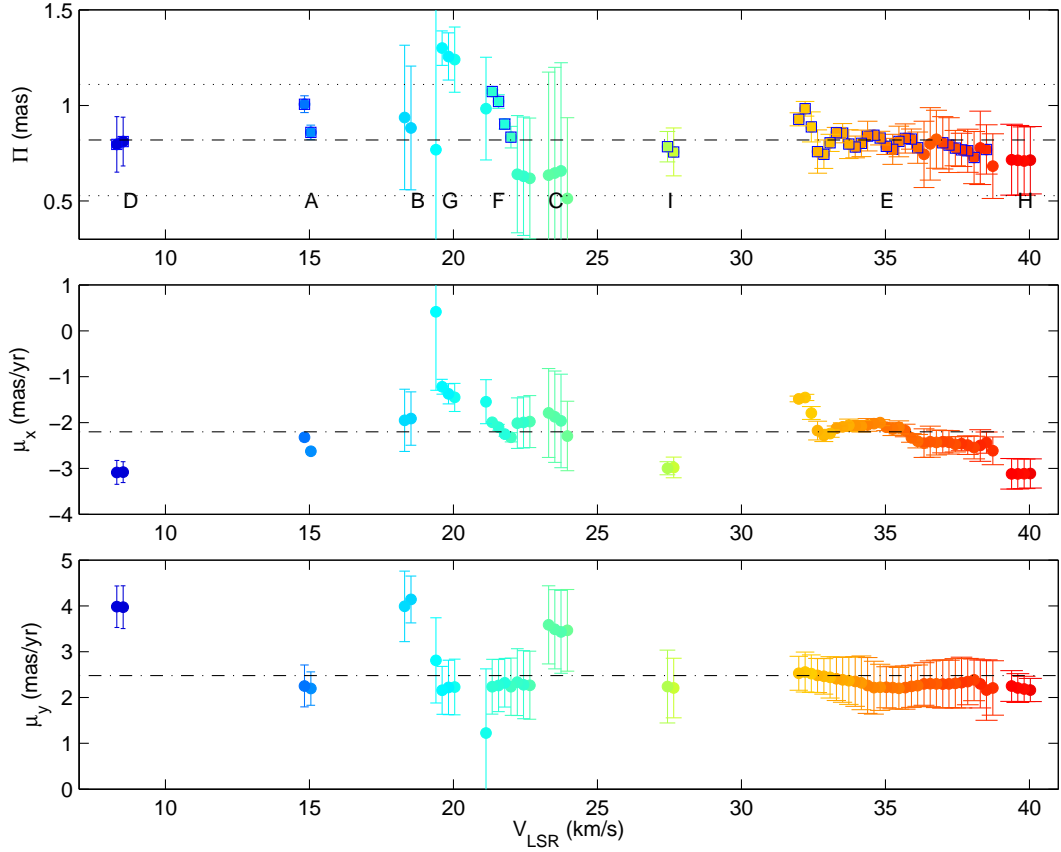}
    \caption{
Individual solutions for parallax and proper motion of all measured
maser spots.  Regions of maser spots are indicated with letters near the
bottom of the {\it top panel}.  The {\it dash-dotted lines} indicate the
means and the {\it dotted lines} indicate the mean $\pm2$ standard
deviations for the parallaxes. 
\newline (A color version of this figure is available in the online journal.)
   \label{fig:para_pm_all}} 
\end{figure}

Fig.~\ref{fig:para_pm_all} shows the estimated parallaxes and proper
motions for all measured maser spots.  While the parallaxes should be
identical (within measurement uncertainties), the proper motions are
expected to vary among the spots owing to internal motions of
$\sim10$~\kms\ ($1.8$ \masy\ at a distance of 1.2 kpc).  While most of
the maser spot parallaxes show good internal consistency, the dispersion
of parallax estimates over the entire ensemble is considerably larger
than the formal errors would suggest.  This is caused by residual
systematic errors affecting the fits, which originate in the complexity
and evolution of blended spectral and spatial structure for some of the
masers.

We discarded all parallax solutions that had formal uncertainties large
than 150 $\mu$as, corresponding to relative uncertainties of $18\%$ (for
a parallax of 0.83 mas).  Some parallaxes from region G appear to be
outliers, and we discarded the results from this region.  The remaining
fits generally yield internally consistent parallaxes.  Most of the
parallax results are distributed in a range defined by $\overline{\Pi}
\pm 2\sigma_{\Pi}$, where $\overline{\Pi}$ and $\sigma_{\Pi}$ are the
average and standard deviation of parallaxes, respectively.
Table~\ref{tab:para_pm_fin} lists the results of the parallax and proper
motion fits for all remaining maser spots. 

Since one expects the same parallax for all maser spots, we did a
combined solution (fitting with a single parallax parameter for all
maser spots, but allowing for different proper motions for each maser
spot) using all the remaining maser spots (indicated with squares in the
top panel of Fig.~\ref{fig:para_pm_all}).  The combined parallax
estimate is $0.830 \pm 0.079$~mas, corresponding to a distance of
$1.20^{+0.10}_{-0.13}$~kpc, which is consistent with the VERA result for
the \hho\ masers in \vycma\ \citep{choi08a}.  The quoted uncertainty is
the formal error multiplied by $\sqrt{n}$ (where $n$ is the number of
maser spots used in the final parallax fit) to allow for the possibility
of correlated position variations for all the maser spots.  This could
result from small variations in the background source or from unmodeled
atmospheric delays, both of which would affect the maser spots nearly
identically~\citep{reid09a}.  The average proper motion of all selected
maser spots is $-2.21\pm 0.06$ \masy\ eastward and $2.29 \pm 0.30$
\masy\ northward, which is reasonably compatible with the VERA results
($-2.21\pm 0.16$ \masy\ eastward and $1.02\pm 0.61$ \masy\ northward).

%--- Fig-  all parallaxes and proper motions
\clearpage
\clearpage

\begin{deluxetable}{crrrrr} 
 \tabletypesize{\scriptsize} %7pt 
% \tabletypesize{\footnotesize} %8pt
\tablecaption{Parallax and proper motion fits\label{tab:para_pm_fin}} 
\tablewidth{0pt} 
\tablehead{ 
\colhead{Region} & \colhead{Ch.} & \colhead{\VLSR}   & \colhead{Parallax} & \colhead{\mux}    & \colhead{\muy}   \\ 
\colhead{}       & \colhead{}    & \colhead{(\kms)}  & \colhead{(mas)}    & \colhead{(\masy)} & \colhead{(\masy)} 
} 
\startdata 
     A & 161  &  15.0  &    0.859 $\pm$  0.038 &   -2.625 $\pm$  0.068 &    2.196 $\pm$  0.365 \\ 
       & 162  &  14.8  &    1.007 $\pm$  0.044 &   -2.323 $\pm$  0.078 &    2.252 $\pm$  0.457 \\ 
     D & 191  &   8.5  &    0.811 $\pm$  0.127 &   -3.081 $\pm$  0.228 &    3.972 $\pm$  0.466 \\ 
       & 192  &   8.3  &    0.796 $\pm$  0.145 &   -3.088 $\pm$  0.260 &    3.982 $\pm$  0.454 \\ 
     E &  53  &  38.5  &    0.770 $\pm$  0.129 &   -2.437 $\pm$  0.231 &    2.164 $\pm$  0.660 \\ 
       &  55  &  38.1  &    0.728 $\pm$  0.137 &   -2.544 $\pm$  0.245 &    2.385 $\pm$  0.455 \\ 
       &  56  &  37.9  &    0.763 $\pm$  0.113 &   -2.492 $\pm$  0.202 &    2.346 $\pm$  0.506 \\ 
       &  57  &  37.6  &    0.768 $\pm$  0.087 &   -2.460 $\pm$  0.155 &    2.325 $\pm$  0.553 \\ 
       &  58  &  37.4  &    0.777 $\pm$  0.093 &   -2.475 $\pm$  0.166 &    2.310 $\pm$  0.538 \\ 
       &  59  &  37.2  &    0.793 $\pm$  0.143 &   -2.431 $\pm$  0.257 &    2.296 $\pm$  0.517 \\ 
       &  60  &  37.0  &    0.805 $\pm$  0.137 &   -2.417 $\pm$  0.246 &    2.296 $\pm$  0.526 \\ 
       &  64  &  36.1  &    0.779 $\pm$  0.082 &   -2.406 $\pm$  0.146 &    2.260 $\pm$  0.453 \\ 
       &  65  &  35.9  &    0.824 $\pm$  0.053 &   -2.334 $\pm$  0.094 &    2.242 $\pm$  0.459 \\ 
       &  66  &  35.7  &    0.827 $\pm$  0.076 &   -2.173 $\pm$  0.136 &    2.217 $\pm$  0.470 \\ 
       &  67  &  35.5  &    0.812 $\pm$  0.080 &   -2.107 $\pm$  0.143 &    2.201 $\pm$  0.456 \\ 
       &  68  &  35.3  &    0.770 $\pm$  0.079 &   -2.142 $\pm$  0.141 &    2.218 $\pm$  0.463 \\ 
       &  69  &  35.0  &    0.786 $\pm$  0.042 &   -2.100 $\pm$  0.075 &    2.225 $\pm$  0.448 \\ 
       &  70  &  34.8  &    0.829 $\pm$  0.036 &   -2.001 $\pm$  0.064 &    2.222 $\pm$  0.503 \\ 
       &  71  &  34.6  &    0.844 $\pm$  0.012 &   -2.025 $\pm$  0.022 &    2.216 $\pm$  0.576 \\ 
       &  72  &  34.4  &    0.840 $\pm$  0.078 &   -2.051 $\pm$  0.139 &    2.262 $\pm$  0.610 \\ 
       &  73  &  34.2  &    0.802 $\pm$  0.070 &   -2.080 $\pm$  0.124 &    2.321 $\pm$  0.585 \\ 
       &  74  &  33.9  &    0.783 $\pm$  0.057 &   -2.071 $\pm$  0.101 &    2.350 $\pm$  0.540 \\ 
       &  75  &  33.7  &    0.798 $\pm$  0.076 &   -2.059 $\pm$  0.135 &    2.366 $\pm$  0.515 \\ 
       &  76  &  33.5  &    0.855 $\pm$  0.050 &   -2.094 $\pm$  0.089 &    2.386 $\pm$  0.488 \\ 
       &  77  &  33.3  &    0.857 $\pm$  0.028 &   -2.123 $\pm$  0.050 &    2.411 $\pm$  0.475 \\ 
       &  78  &  33.1  &    0.804 $\pm$  0.045 &   -2.240 $\pm$  0.080 &    2.442 $\pm$  0.452 \\ 
       &  79  &  32.9  &    0.745 $\pm$  0.074 &   -2.280 $\pm$  0.133 &    2.462 $\pm$  0.388 \\ 
       &  80  &  32.6  &    0.759 $\pm$  0.113 &   -2.175 $\pm$  0.203 &    2.481 $\pm$  0.373 \\ 
       &  81  &  32.4  &    0.888 $\pm$  0.078 &   -1.792 $\pm$  0.140 &    2.523 $\pm$  0.409 \\ 
       &  82  &  32.2  &    0.983 $\pm$  0.038 &   -1.453 $\pm$  0.068 &    2.557 $\pm$  0.436 \\ 
       &  83  &  32.0  &    0.928 $\pm$  0.034 &   -1.485 $\pm$  0.061 &    2.529 $\pm$  0.369 \\ 
     F & 129  &  22.0  &    0.834 $\pm$  0.057 &   -2.322 $\pm$  0.101 &    2.236 $\pm$  0.623 \\ 
       & 130  &  21.8  &    0.903 $\pm$  0.027 &   -2.252 $\pm$  0.048 &    2.322 $\pm$  0.530 \\ 
       & 131  &  21.6  &    1.021 $\pm$  0.036 &   -2.102 $\pm$  0.064 &    2.261 $\pm$  0.568 \\ 
       & 132  &  21.3  &    1.073 $\pm$  0.009 &   -1.993 $\pm$  0.016 &    2.237 $\pm$  0.598 \\ 
     I & 103  &  27.6  &    0.757 $\pm$  0.125 &   -2.979 $\pm$  0.224 &    2.208 $\pm$  0.649 \\ 
       & 104  &  27.4  &    0.784 $\pm$  0.079 &   -2.997 $\pm$  0.142 &    2.240 $\pm$  0.794 \\ 
%\hline
       &      &        &                       &                       &                       \\ 
Combined &    &        &    0.830 $\pm$  0.079 &   -2.210 $\pm$  0.060 &    2.290 $\pm$  0.300 \\ 

\enddata 
\small{
Note. --- Absolute proper motions are defined as $\mux = \mu_{\alpha \cos{\delta}}$ and $\muy = \mu_{\delta}$.
}
\end{deluxetable} 
\clearpage

\section{ABSOLUTE POSITION OF CENTRAL STAR}
\label{sec:abs_pos}

\subsection{Absolute position of \vycma\ derived from VLBA observation}
\label{subsec:vlba_pos}

Based on the circumstellar maser distribution, it is possible to locate
the central star with reasonable accuracy.  For example, for a ring-like
distribution of maser emission, the star should be located at the center
of the ring.  That this is indeed the case has been observationally
verified for AGB stars by \citet{reid90} for \hho\ masers and
\citet{reid07} for SiO masers.  Previous VLBI observations of SiO
masers towards \vycma\ have shown that the masers appear clumpy and
asymmetrical \citep{miyoshi94, miyoshi03}.  Our observations (see
Fig.~\ref{fig:vycma_maser_pos}) show that the SiO masers toward \vycma\
exhibit only a partial ring-like structure.

\begin{figure}[H]
  \centering
  \includegraphics[angle=0,scale=0.95]{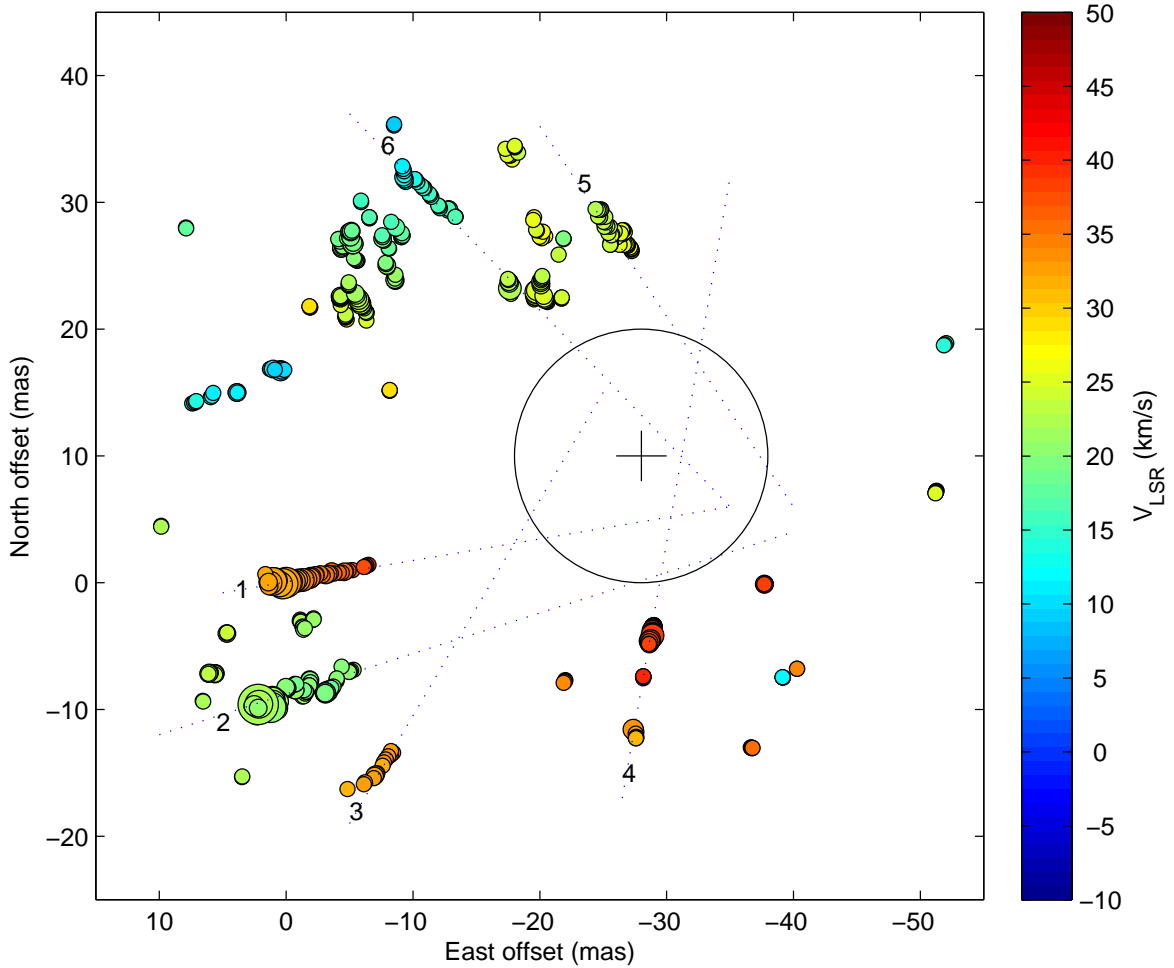}
  \caption{
Positions of SiO masers for 2006 April 16 and inferred position of the
central star ({\it cross}) with a radius ({\it circle}) of 10 mas
estimated from mid-infrared observations~\citep{monnier00}, Each maser
spot is represented by a {\it filled circle} whose area is proportional
to the logarithm of the flux density.  The {\it dotted lines} labeled
with a number from ``spoke-like'' maser features point back to the
central star.  The \VLSR\ velocity of the maser spots is color-coded as
indicated by the color bar on the right side. 
\newline (A color version of this figure is available in the online journal.)
}
  \label{fig:circ} 
\end{figure}

There are ``spoke-like'' (linearly distributed) maser features at all
four epochs; one of the best examples is found in the group of features
which include the reference spot (see Fig.~\ref{fig:circ}).  These maser
features, appear to be composed of ``spokes" of gas flowing outward from
the central star and mostly have decreasing radial velocities with
increasing the distance from the star~\citep{yi05, matsumoto08}.
There are several spoke-like maser features in \vycma\ and they probably
``point back'' to the star.  From these spoke-like features and the
ring-like maser distribution, we estimate the position of the central
star relative to the reference spot to be (\dx\ = $-$28, \dy\ = $+$10
mas) in east and north directions with an uncertainty of about 10 mas.
Based on the combined parallax fit in \S\ref{sec:para}, we determine the
position offset (\dx\ = $-4.8 \pm 0.1$, \dy\ = $1.6 \pm 0.2$ mas) of the
reference spot at mean epoch 2006.53 of our observation.  Combining
these offsets, we estimate the absolute position (\ra\ =
07\h22\m58\decs3259, \dec\ = $-$25\deg46\arcmin03\decas063) of the
central star with an uncertainty of about 10 mas.  

\clearpage

\subsection{SiO masers and radio photosphere from VLA observation at 43
GHz}
\label{s1sec:43vla}

From multi-wavelength VLA observations of a sample of long period Mira
and semi-regular variable stars, \citet{reid97,reid07} detected
emission from their radio photospheres.  The approach of using a strong
narrowband signal (maser emission) in one observing band to calibrate
the phase (and amplitude if necessary) of a weak broadband signal (radio
continuum emission from the radio photosphere) was described in detail
by \citet{reid97,reid07}.  In the following we refer to the two
datasets as narrow band (NBD) and broad band (BBD) data.

In order to register the SiO masers relative to the radio photosphere,
we first measured the position of the radio photosphere from the BBD
relative to the maser emission in the (pseudo-continuum) NBD.  Next, we
aligned the maser emission from the spectral-line data by producing a
map of ``simulated" NBD (hereafter SNBD) from the line data.  Because
the NBD and SNBD cover the same velocity range, we can align the maps
from the NBD and SNBD by comparison of positions of peaks in each map.
This allows the positions of the emission in individual channels to be
registered to the NBD, and then to the BBD (radio photosphere) emission.

Following the procedures described in detail in \citet{reid97}, once
the data were ``cross-self-calibrated", both NBD and BBD were imaged
with the AIPS task IMAGR.  We performed Gaussian position fits to the
peak in the BBD and NBD, using AIPS task JMFIT, and obtained the
relative position (\dx\ = $-16 \pm 2$, \dy\ = $-6 \pm 2$ mas). 

The line data were self-calibrated by choosing a channel with strong
emission as a reference, and the resulting phase and amplitude
corrections were applied to the other channels.  From the line data we
produced a spectral-line data cube and then we produced a NBD-like map
by averaging all the channels in the cube.  We fitted the two strongest
peaks in the NBD and SNBD maps and calculated a relative position of
(\dx\ = $-$8, \dy\ = $+$24 mas) with an uncertainty less than 1 mas.

\begin{figure}[H]
  \centering
  \includegraphics[scale=0.8,angle=-0]{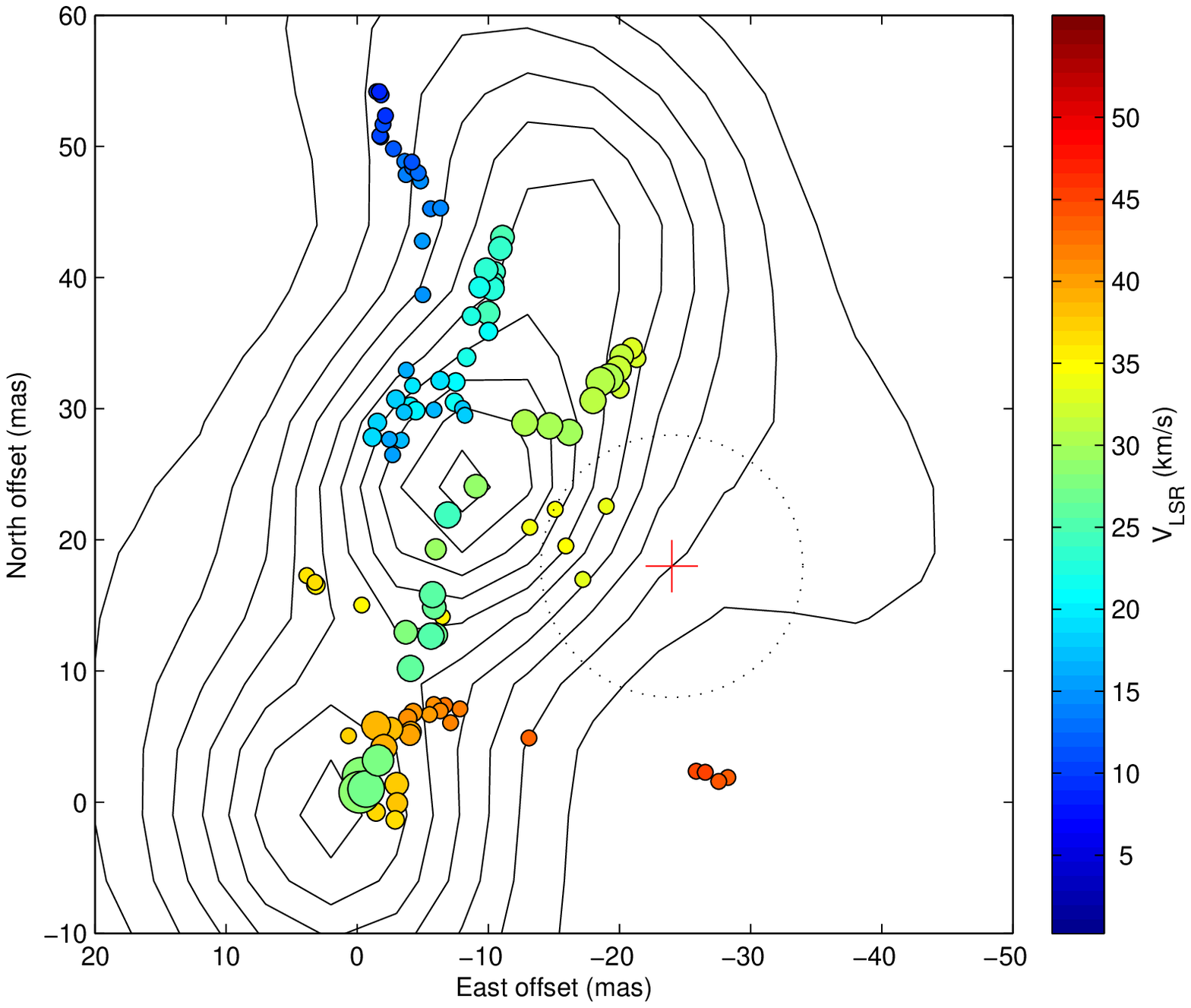}
  \caption{
SiO maser distribution around \vycma. All offsets are relative to the
best fit position of the peak in the map of the strongest maser emission
at \VLSR\ of 22.1 \kms.  All symbols are overlaid on contours of the 20
mas resolution image of the integrated narrowband emission (SNBD).
Contour levels are spaced linearly at 5 times the rms noise level of
0.69 \jybeam.  All the maser spots are marked with {\it circles}, whose
sizes are proportional to the logarithm of their flux densities.  The
best fit position of the radio photosphere (BBD) is marked as a {\it
cross}.  The {\it dotted circle} presents the boundary of the central
star with a radius of 10 mas from \citet{monnier00}.  The \VLSR\ of the
maser spots is color-coded as indicated by the color bar to the right.
\newline (A color version of this figure is available in the online journal.)
}
  \label{fig:maser_bbd}
\end{figure}

Finally, we estimated a position offset (\dx\ = $-24$, \dy\ = $+18$ mas)
of the weak continuum source (radio photosphere) relative to the maser
spots (reference spot) with a uncertainty of 3 mas. Thus, all the maser
spots were registered to the radio photosphere at a similar accuracy
(see Fig.~\ref{fig:maser_bbd}).

However, to estimate {\it absolute} position of the continuum source, we
have to align the maser spots with those from the VLBA observations.
Due to the significantly different angular resolution and sensitivity of
the VLA and VLBA, there is no guarantee that the two images should be
identical.  Note that the VLA has a 50 mas synthesized beam, which for
any channel essentially gives a centroid position.  If a single spectral
channel contains emission from across the source, the VLA position would
be falsely interpreted as near the center of the distribution.

\begin{figure}[H]
  \centering
  \includegraphics[scale=0.9,angle=-0]{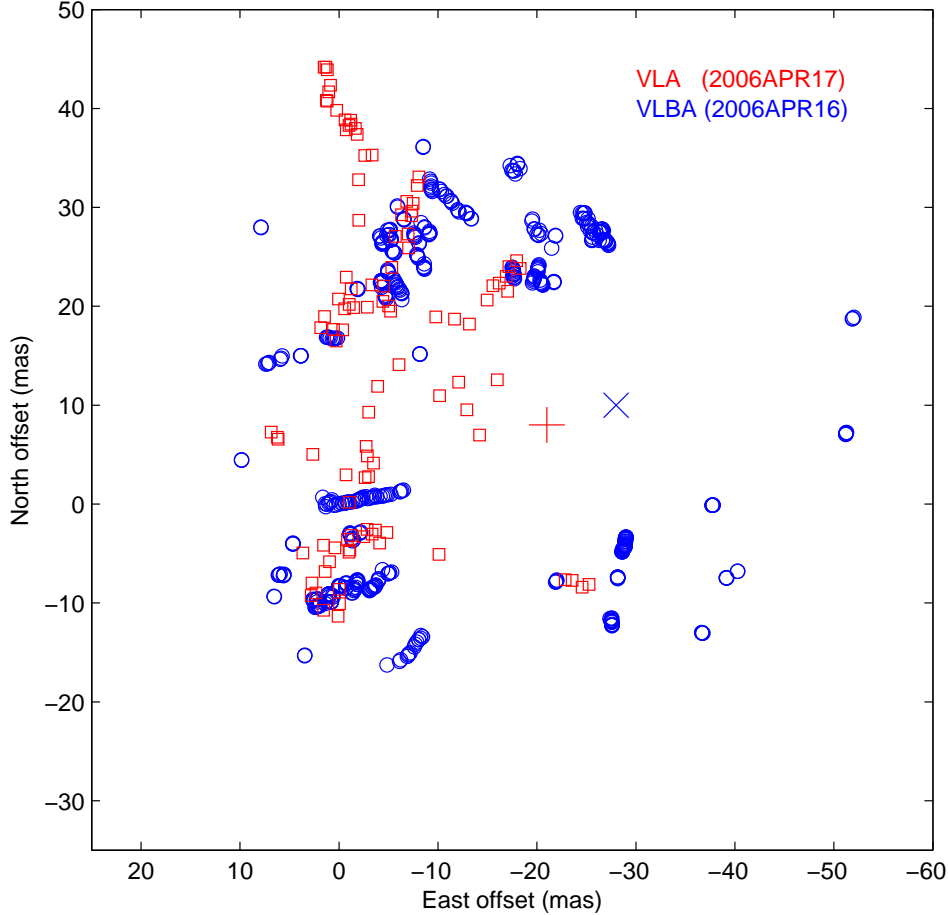}
  \caption{
Comparison of SiO maser distributions from the VLBA ({\it circle}) and
VLA ({\it square}), after cross registration (see text). The observation
epochs are also labeled.  The position of the star determined from the
VLA radio photosphere observations is indicated by the ``$+$" and from
the VLBA ``spoke-like'' maser spot distribution by the ``$\times$"
symbols.  
\newline (A color version of this figure is available in the online journal.)
}
  \label{fig:cf_vlba_vla}
\end{figure}

Nevertheless, the SiO maser distributions from VLA and VLBA data bear
significant resemblance.  In order to align these two maps, we assumed
that the centroid positions of the strongest maser spots are consistent.
For the VLA map, the strongest maser spot was chosen as the phase
reference and, hence, is at the map origin (see
Fig.~\ref{fig:maser_bbd}).  For the VLBA map, the phase reference was
not the strongest spot, and the reference spot is at ($+3,-10$) mas (see
Fig.~\ref{fig:circ}).  After shifting all positions from the VLA map by
this offset, we find reasonable agreement with the VLBA and VLA maser
spot maps.  Since we obtained an offset of ($-7,+2$) mas for the central
star (imaged with the VLA) relative to the VLBA spots maps, we now can
locate the star on the VLBA spot maps to about 10 mas uncertainty.  This
is shown in Fig.~\ref{fig:cf_vlba_vla}.

\section{THE KINEMATIC MODEL}
\label{sec:kin_mod}

\subsection{Expansion}
\label{subsec:exp_mod}

Combining the proper motion (\mux, \muy) and parallax with the \VLSR\ of
each maser spot, we can derive its 3-dimensional velocity vector
$V_i(v_{ix},v_{iy},v_{iz})$ and its variance  
\begin{eqnarray}
&&  \sigma_{ix}^2 = (k\sigma_{\mu ix}) ^2 + v_{turb}^2 \\
&&  \sigma_{iy}^2 = (k\sigma_{\mu iy}) ^2 + v_{turb}^2  \\
&&  \sigma_{iz}^2 = \sigma_{V_{lsr}}^2 + v_{turb}^2 
\end{eqnarray}
where $k$ is used to convert proper motion to velocity adopting the
distance of 1.2 kpc from \S\ref{sec:para}, and $v_{turb}$ is a
``turbulent'' velocity component, assumed to be $\sim 3$ \kms.  The
\VLSR\ of the star (\VLSRS) is adopted as 22 \kms\ from
\citet{menten06}. 

The model velocity vector for an expanding flow is \begin{equation}
  \hat{{V}}_i = V_0 +  v_{exp} \frac{r_i}{|r_i|} \label{eq:model}
\end{equation}
where, $v_{exp}$ is an expansion speed modeled as
\begin{equation}
  v_{exp} = v_{exp0} + a |r_i|~~,
\end{equation}
$V_0(v_{0x}, v_{0y}, v_{0z})$ is the systemic velocity (presumably of
the exciting star) and $r_i$($R_i-R_0$) is the position vector of a
maser spot $R_i(x_i,y_i,z_i)$ relative to the expansion center
$R_0(x_0,y_0,z_0)$.  The sky positions ($x_i,y_i$) of the maser spots
are relative to the reference maser spot which is located at
($x=0$,$y=0$), and the radial position $z$ is relative to the expansion
center assuming $z=0$.  The sky positions ($x, y$) are converted from
mas to AU adopting a distance of 1.2 kpc.  

The three components of velocity for all maser features in
\S\ref{sec:int_mot} are taken as observables, and the weight of each
observable is assigned the reciprocal of its variance.  The fitting
procedure is similar to that used in previous
studies~\citep{reid88,gwinn92,imai00}; however, unlike most other
expanding maser models, we know the distance and $v_{0z}$ and do not
have to solve for them.  Apart from the global parameters listed in
Table~\ref{tab:exp_par},  the $z_i$ of each maser spot is also solved
for as a free parameter.  In order to restrict the parameter estimates
to physically reasonable values as well as to incorporate our knowledge
of parameters $x_0$ and $y_0$, we added {\it a priori} information for
these two parameters by adding extra data-like equations, one for each
parameter, to the observables, and weighting the ``pseudo-observables''
with their estimated uncertainties.  This allowed us to incorporate the
information that ($x_0 = -33 \pm 12, y_0 = 12 \pm 12$ AU) from
\S\ref{sec:abs_pos}. 

\begin{figure}[H]
  \centering
  \includegraphics[scale=0.85]{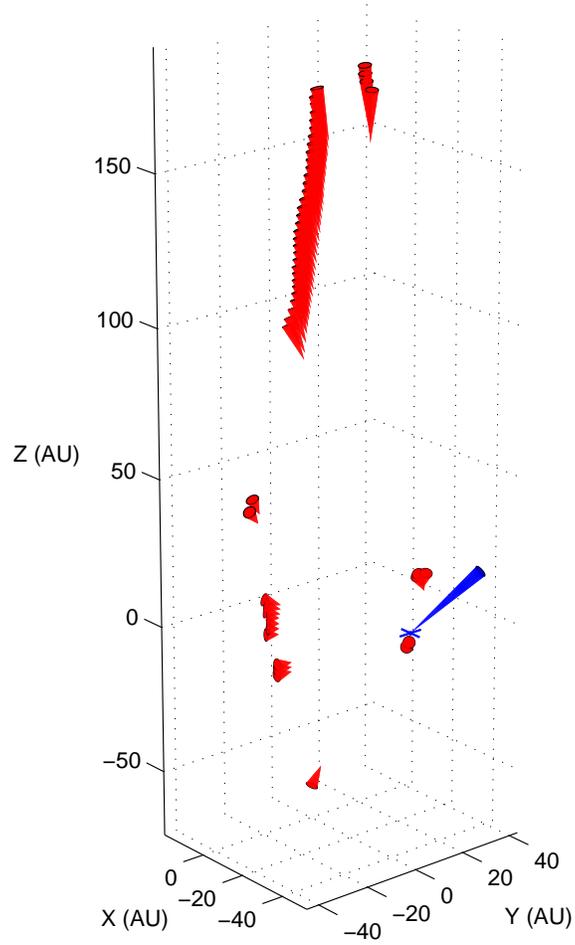}
  \caption{
Positions and velocities for maser spots. The maser spots are located at
the apexes of the {\it red cones}.  The {\it red cone} lengths and
orientations show the speed and direction of their velocities after
removal of the systemic velocity $V_0$, which is indicated with {\it
blue cone}. The expansion center is located at the apex of the blue cone
with the indicated error bar in X and Y directions.
\newline (A color version of this figure is available in the online journal.)
}
  \label{fig:exp}
\end{figure}

The global parameters and the $z_i$ of each maser spot were adjusted to
minimize the sum of the squares of the weighted residuals using the
Levenberg-Marquardt method.  We first performed the model fitting using
all the maser spots, and then discarded those maser spots (6 of 60) with
large residuals ($> 3\sigma$), and estimated the parameters again; the
best fitted global parameters are listed in Table~\ref{tab:exp_par}.  As
Fig.~\ref{fig:exp} shows, the expansion velocities are much smaller than
the systemic velocity, indicating that the absolute proper motions of
the maser spots are dominated by the systemic velocity of $-16 \pm 1$
\kms\ eastward and $15 \pm 1$ \kms\ northward (corresponding to $\mux =
-2.8 \pm 0.2$ \masy\ and  $\muy = 2.6 \pm 0.2$ \masy), which is
consistent with the proper motion estimated in \S\ref{sec:para} within
3$\sigma$.

\begin{table}[H]
  \centering
  \caption[]{Kinematic Model of \vycma\ SiO masers}
  \smallskip
  \begin{tabular}{crlll}
  \hline\hline
Parameter & Value &  Uncertainty & Units & Comments\\
\hline
%--- with a
$x_{0}$     & $-$35 & $\pm$ 5 & (AU)   & Center of expansion\\
$y_{0}$     &    13 & $\pm$ 5 & (AU)   & \\
$v_{0x}$    & $-$16 & $\pm$ 1 & (\kms) & Systemic velocity \\
$v_{0y}$    &    15 & $\pm$ 1 & (\kms) & \\
$v_{exp0}$  &     2 & $\pm$ 1 & (\kms) & Expansion speed at the origin\\
$a$         &     0.09 & $\pm$  0.01 & (\kms~AU$^{-1}$) & Expansion speed gradient \\
 \hline
  \end{tabular}
  \label{tab:exp_par}
\end{table}

\clearpage
\subsection{Radial Spoke-like Maser Features}
\label{subsec:spoke}

%--- Fig- spoke vrad vs distance
\begin{figure}[H]
  \centering
  \includegraphics[angle=0,scale=0.80]{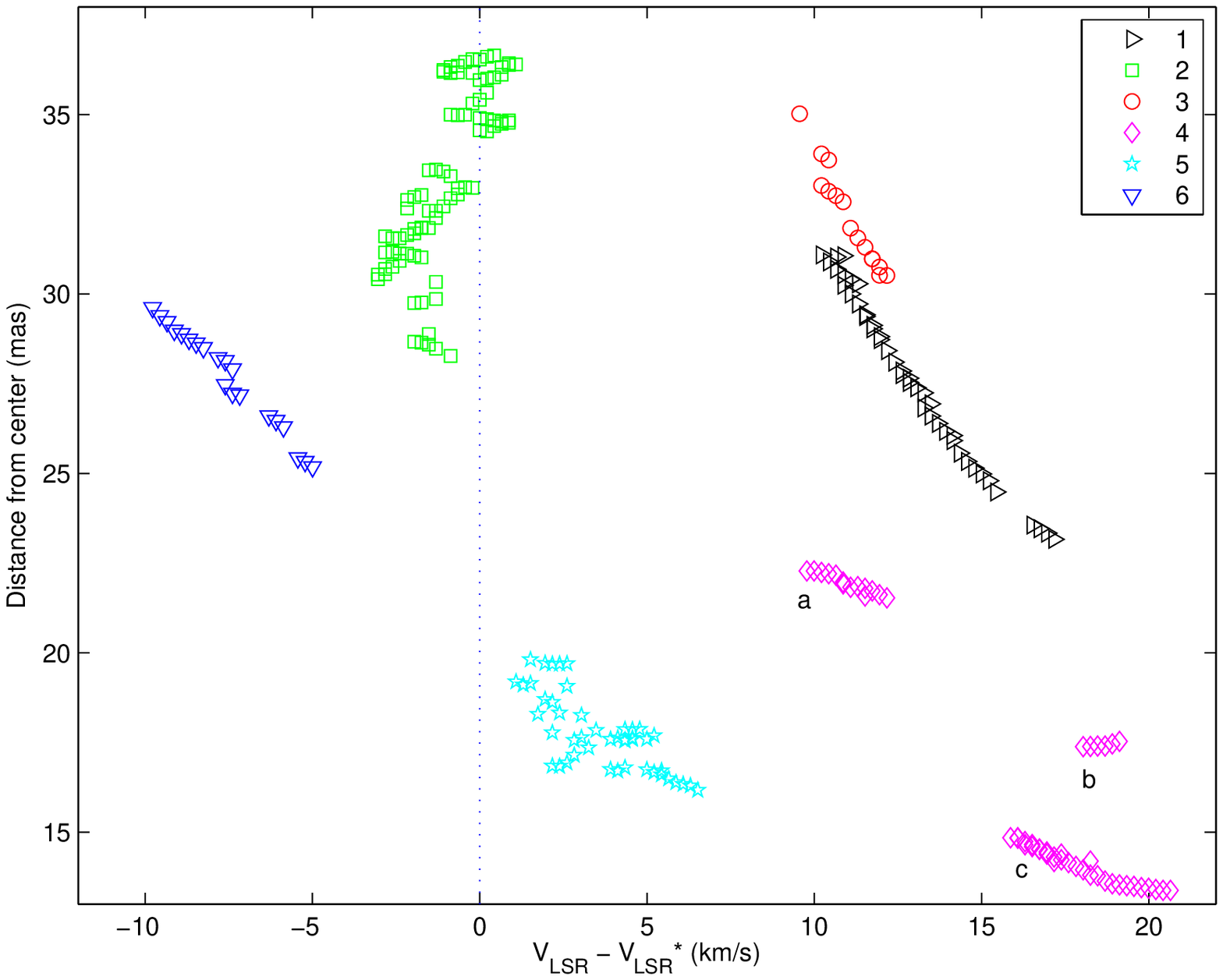}
  \caption{
Angular distances versus radial velocities (\VLSR $-$ \VLSRS) of
spoke-like maser features. Different {\it markers} denotes different
maser features from Fig.~\ref{fig:circ}. Feature 4 is divided into three
parts (a, b and c). 
\newline (A color version of this figure is available in the online journal.)
} 
  \label{fig:dist_vel} 
\end{figure} 

As mentioned in \S\ref{subsec:vlba_pos}, there are several spoke-like
maser features in \vycma.  Fig.~\ref{fig:dist_vel} shows the
distance-velocity structure of six of these features, labeled as in
Fig.~\ref{fig:circ}.  These spoke-like features have clear velocity
gradients, and most of them have radial velocities (\VLSR\ $-$ \VLSRS)
$>$ 0 and decreasing \VLSR\ with increasing angular distance from the
center.  This could happen in two ways: (1) the spoke-like features have
motions outward from the star (i.e., outflow) on the far side and are
decelerating; (2) the spoke-like features have motions inward toward the
star (i.e., infall) on the near side and are accelerating.  Taken into
consideration the motions shown in Fig.3, we believe most of the
spoke-like features are on the far side and are decelerating.

In order to explore the nature of the radial velocities of the observed
spoke-like features, we adopt the ballistic-orbit model
of~\citet{matsumoto08} to fit the distance-velocity structure of the
spoke-like feature.  This model assumes that the maser spots are moving
in a radial direction with acceleration or deceleration due to the
stellar gravity. 

The relation between the projected distance from the center of the star,
$r$, and the line-of-sight velocity, $v$, is written as
\begin{equation} 
r = \frac{r_0 \cos{i} } {1+ \dfrac{r_0}{2GM}\left(\dfrac{v}{\sin{i}}\right)^2 }~~,
\label{eq:spoke}
\end{equation}
where $v$ = \VLSR\ $-$ \VLSRS, $G$ is the gravitational constant, $M$
the stellar mass, $i$ ($\neq 0$) the inclination angle of the maser
feature projected onto the sky plane, and $r_0$ is the distance of
apocenter (where $v=0$) from the center of the star.  However, unlike
the ballistic-orbit model fitting in~\citet{matsumoto08}, we adopt a
stellar mass of 25 \msun\ from~\citet{monnier00}, while we think $r_0$
is unknown and may be different for each spoke-like feature. Thus, we
solve for $r_0$, which is converted to angular units adopting a distance
of 1.2 kpc, and $i$ for each spoke-like feature.

%--- Fig- spoke model fitting 
\begin{figure}[H]
  \centering
  \includegraphics[angle=0,scale=0.80]{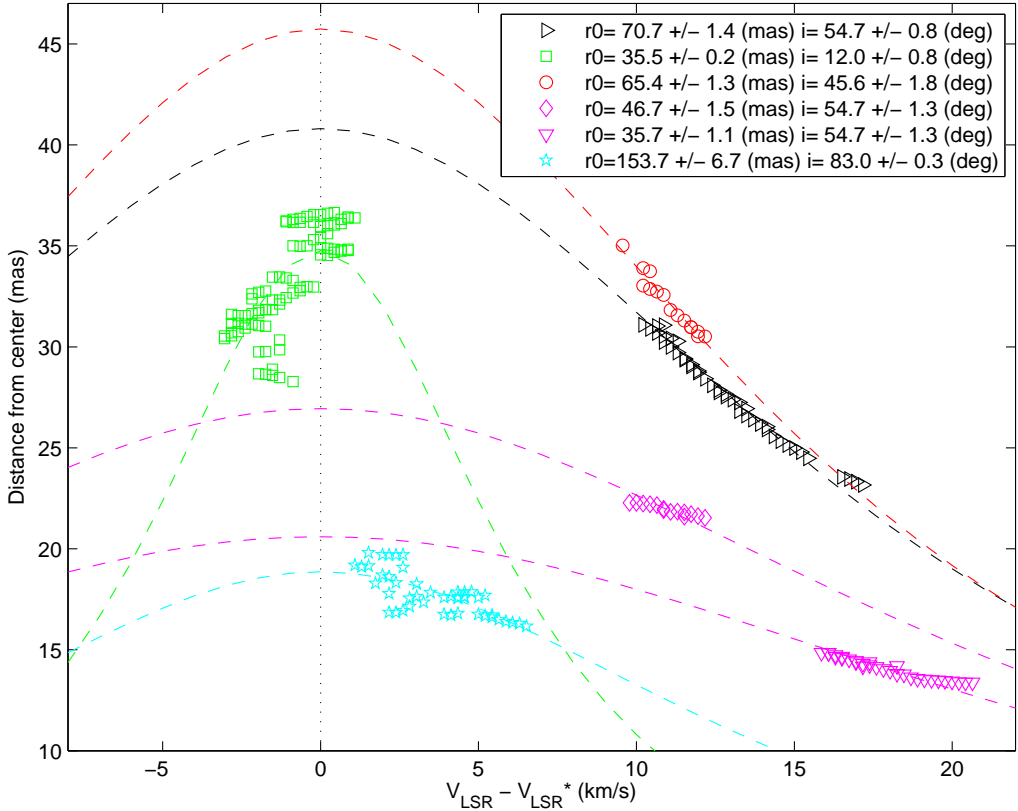}
  \caption{
Fitted ballistic orbit ({\it dashed lines}) for the spoke-like feature
({\it markers}) in the distance-velocity diagram.  Different {\it
markers} denote different maser features labeled in
Fig.~\ref{fig:dist_vel}. The fitted parameters and their uncertainties
are labeled in the top right corner.
\newline (A color version of this figure is available in the online journal.)
} 
  \label{fig:spoke_fit} 
\end{figure}

Fig.~\ref{fig:spoke_fit} shows the observed and modeled
distance-velocity structures of all the spoke-like features (except 6
and 4b which do not follow a ballistic model) using Eq.~\ref{eq:spoke}.
The distance-velocity structure should be symmetric about $v=0$, which
means blue shifted ($v < 0$) maser feature should increase their \VLSR\
with increasing distance.  Since most of the spoke-like features can be
reasonably well modeled with ballistic orbits, this suggests that most
of the spoke-like features ``point back'' to the star.

\clearpage

\section{DISCUSSION}
\label{sec:disc}

% maser distribution and kinematics 

Images of some AGB stars such as IK Tauri demonstrate that SiO masers
may form circular or elliptical rings of emission.  Also the
line-of-sight velocity structure of some SiO masers can have an apparent
axis of symmetry consistent with the elongation axis of the maser
distribution, suggestive of rotation of the SiO maser shell \citep[and
references therein]{boboltz05}.  However, for \vycma\ such symmetry is
not evident in the spatial or velocity distribution of the masers.
Large scale optical and near-infrared images of \vycma\ (e.g.
\citet{kastner98, monnier99, smith01, humphreys07}) display a possible
northeast-southwest axis of symmetry, which might suggest a possible
bipolar structure. However, these observations are from dust emission
much further from the star than the circumstellar SiO masers.  After
subtracting the average internal motion,  the SiO masers show a tendency
for expansion from the central star, which is similar to the 22 GHz
\hho\ masers measured by VERA~\citep{choi08b}.  Observation of SiO ($v =
0, J = 1 \to 0$) maser emission towards \vycma\ with VLA shows a bipolar
outflow along the line of sight \citep{shinnaga04}, consistent with the
3-dimensional kinematics of the \hho\ masers from~\citet{choi08b}.
However, there is no strong evidence for bipolar outflow but a slow
quasi-spherical outflow displayed in Fig.~\ref{fig:exp} from our VLBA
observations.

% expansion and spoke feature

The 3-dimensional expansion model implies that the radial velocities of
the maser features will increase with increasing distance from the star.
If the maser features are radially aligned as the ballistic orbit model
assumes, then the expansion model seems counter to the ballistic model.
However, the ballistic-orbit model provides only one of the possible
explanations for the spoke-like features. As mentioned in
\S\ref{subsec:spoke}, there are still spoke-like features that cannot be
modeled with a ballistic orbit.  A more realistic model might involve a
hydrostatic inner envelope and/or acceleration driven by pulsation or
giant convective cells first introduced by~\citet{schwarzschild75}.  For
the expansion model, a very small value of the expansion speed gradient
(see Table~\ref{tab:exp_par}) is rather weak evidence for acceleration
in outward motion. Furthermore, even though we know the distance of the
star, based on the combined parallax of the maser spots, the distances
of the maser features from the star are unclear (i.e., coordinate
component along the line of sight, $z$) and can only be solved for as
parameters, which are dependent on the model.  Nevertheless, the
uncertain $z$'s of the maser features for the expansion model will not
significantly affect the derivation of the proper motion of the central
star, which is one of the most important results in this paper.

% parallax and proper motion

Our accurate parallax based on SiO masers measured with VLBA agrees with
that determined from \hho\ masers with VERA (see
Table~\ref{tab:src_abspos}).  The small difference in proper motions of
the center of expansions for the two maser species is not strongly
statistically significant and, even if real, might be caused by small
errors in determining the center of expansion for the different maser
species which are located in different places in the circumstellar
envelope. 

% Compared with main sequence fitting and photometric distance

\citet{lada78} suggested that \vycma\ and the open cluster NGC 2362 are
at nearly the same distance, which was estimated to be 1.5 kpc from
main-sequence fitting of NGC 2362.  Their argument rested on the fact
that \vycma\ was located at the apex of an arc of bright emission that
is visible in red-light images and had a \VLSR\ similar to that of the
molecular cloud complex and the stars of the cluster NGC 2362.  Lada \&
Reid also pointed out that there were other possible ionizing sources
for the arc, including $\tau$ CMa (HD 57061); UW CMa (HD 57060, 29 CMa);
and NN CMa (HD 58011).  These stars were estimated to be 1.1 kpc
distant, based on photometric measurements, and significantly closer
than NGC 2362.  Since our trigonometric parallax is closer to 1.1 than
1.5 kpc, this suggests that the main-sequence fitting at the time was
less accurate than the photometric distances. 

The main-sequence fitting distances are nearly 40 years old and one
might worry that better calibrations would resolve the discrepancy.
However, more recent estimates of the distance to NGC 2362 are 1.62 kpc
\citep{mermilliod86}, 1.42 kpc \citep{brown86}, 1.48 kpc
\citep{moitinho01} and 1.36 kpc \citep{mayne08}.  These are all
reasonably consistent with the 1.5 kpc value previously used and still
different from the parallax values determined in this paper and by VERA.
For an individual star, a photometric distance is highly dependent on
the spectral type and extinction model.  Compared to 1.1 kpc quoted by
\citet{lada78}, \citet{kaltcheva00} estimate a significant different
photometric distance of 2.86 kpc for NN CMa and 1.51 kpc for UW CMa.
These are significantly larger than 1.2 kpc distance to \vycma\ reported
in this paper.  Provided that \vycma\ is indeed associated with NGC 2362
and NN CMa or UV CMa, this suggest that systematic uncertainties for the
distance estimated using either main-sequence fitting or the photometric
method is at least $\sim20\%$.

The absolute position of the central star from our VLBA and VLA
measurement at 43 GHz is consistent with that from VERA.  However, the
position and proper motion from Hipparcos are significantly different
with the values determined from radio measurements (see
Table~\ref{tab:src_abspos}).  Using the position and proper motion from
radio measurement, we predicted the position of VY CMa at the same epoch
as the Hipparcos catalogue, then compared the position derived from
optical and radio observations, and found a large discrepancy of tens of
mas.  \vycma\ is embedded in an asymmetric dust reflection nebula, and
recent images from optical \citep{kastner98, smith01, humphreys07}
and near-IR observations \citep{monnier99} show that the dust shell is
time variable.  Since most of the optical radiation coming from the star
is reprocessed by the surrounding dust, a likely explanation for the
difference between Hipparcos and the radio results is that the observed
optical light is mostly scattered by circumstellar dust and a reliable
stellar proper motion determination at optical wavelengths is
practically impossible. 

Another possibility for a difference between the VLBA and Hipparcos
proper motions would be that an inhomogeneity on \vycma's surface
contributes to the Hipparcos proper motion.  For example, the early
M-type supergiant $\alpha$ Orionis' ultraviolet appearance, as resolved
with the Hubble space telescope, shows a conspicuous, compact ``hot
spot'' \citep{gillilant96}, with a $>$ 200 K higher temperature than the
rest of the surface.  One of their interpretations for this spot is in
the context of a conjecture by \citet{schwarzschild75} that a red giant
star's surface only may show a few hundred convective cells, with large
temperature differences (of up to 1000 K) between the hot and cool
portions of a convective element. In addition, \citet{soker99}
investigated the structure of cool magnetic spots in the photospheres of
evolved stars and found the spots will cause the AGB star to appear
asymmetrical.  Adopting a diameter of 20 mas \citep{monnier00} for
\vycma, could the apparent proper motion from Hipparcos be the result of
a hot spot rotating with the star's surface?  Unfortunately, the
rotation velocities of red supergiant are expected to be of order 1
\kms, much lower the measured proper motion discrepancy.  Could the
random appearance and disappearance of different independent spots mimic
the Hipparcos proper motion?  Given that Hipparcos' proper motion
determinations are typically based on 30 measurements within the mission
lifetime of 3 yr, this is conceivable.

\begin{figure}[H]
  \begin{center}
    \includegraphics[angle=0,scale=0.75]{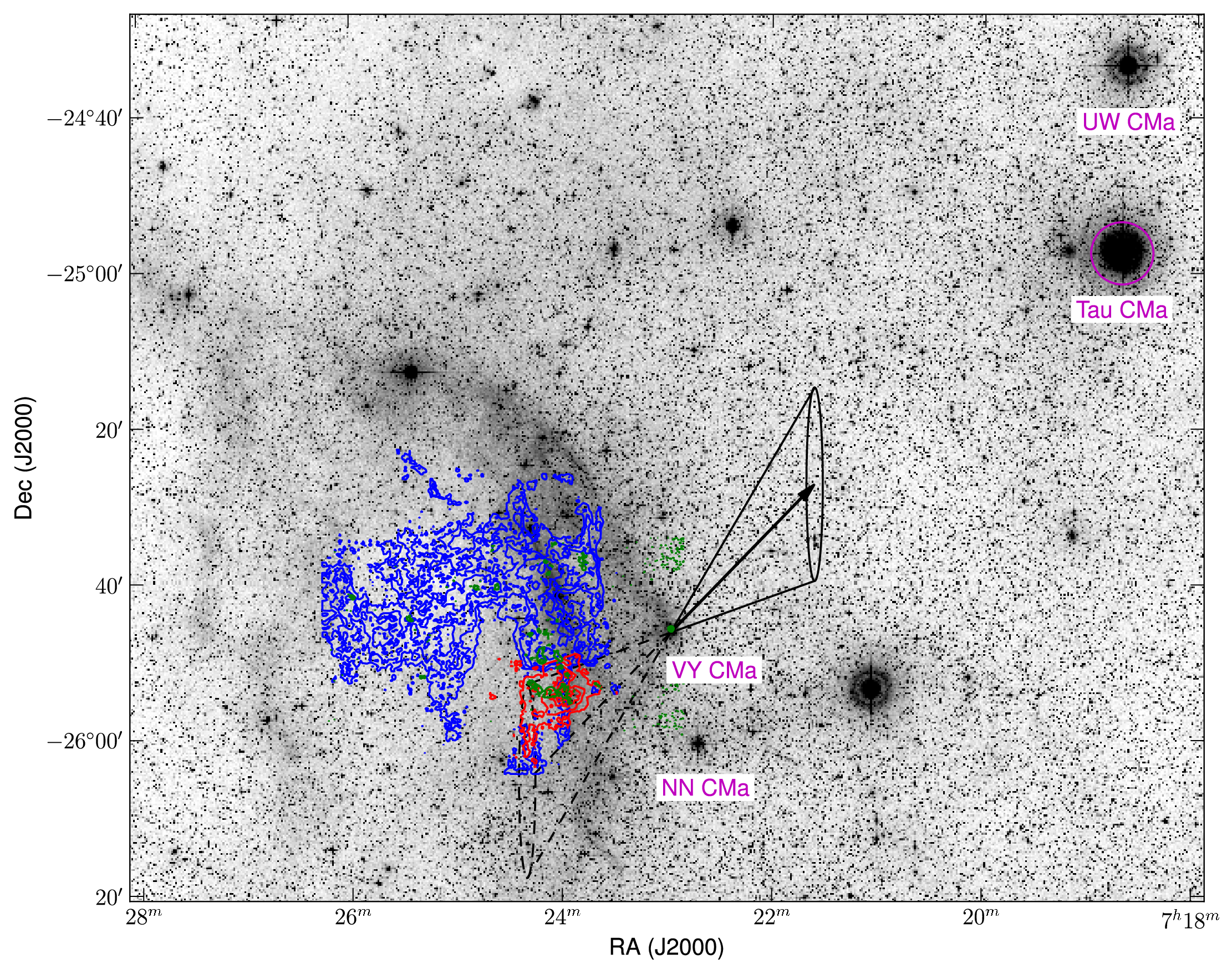}
  \end{center}
  \caption{
H$\alpha$ emission ({\it gray scale}) superposed on the radio continuum
emission at 870 $\mu$m with the Large Apex BOlometer CAmera (LABOCA)
({\it green contours}), and the CO(3-2) emission line is integrated over
15-21 \kms\ ({\it red contours}) and over 21-27 \kms\ ({\it blue
contours}).  \vycma\ is located at the bright rim of the V-shaped
structure of the cloud complex and also a point source in the continuum
emission contours.  The {\it arrow} originating at \vycma\ represents a
motion over 0.5 Myr adopting the absolute proper motion derived in this
paper.  The motion uncertainty is indicated as an {\it error ellipse}
with a confidential level of 95\% centered on the head of the {\it
arrow}; the {\it dashed error ellipse} denotes the probable original
position of \vycma\ 0.5 Myr ago.  Three candidates for the sources of
ionization in this region: Tau CMa, UW CMa and NN CMa from
\citet{lada78} are labeled.  The apparent size of $\sim$ 8\arcmin\ of
the nearby O-star cluster NGC 2362 centered on Tau CMa is indicated as a
{\it circle}.
\newline (A color version of this figure is available in the online journal.)
  }
  \label{fig:vyc}
\end{figure}

%--- Possibility of the proper motion of VY CMa
Fig.~\ref{fig:vyc} shows the geometry of \vycma's optical environment
overlaid with radio emission contours measured with the Atacama
Pathfinder Experiment (APEX) telescope by \citet{schuller11}.  The
V-shaped surrounding material highlighted as H$\alpha$ emission, which
is either external to \vycma\ (i.e., interstellar material) or comes
from \vycma\ (stellar mass loss), suggests relative motion of gas and
the star. Taking our VLBA absolute proper motion, one explanation could
be that \vycma\ is drifting out of the molecular cloud to the east of
it.
%possibly owing to galactic shocks from spiral arms or magnetic forces,
%which are potentially dominant sources of turbulence in the interstellar
%medium, acting on gas.

Assuming that \vycma\ was formed in the region where current signs of
star formation are found (i.e., to the south-east of the molecular cloud
as indicated by the color contours in Fig.~\ref{fig:vyc}) and that the
absolute proper motion of the molecular cloud is small, then the travel
time of \vycma\ to its current location is about 0.5 Myr.  This might be
reasonable since \vycma\ is a very massive, and hence short-lived, star.
However, adopting a luminosity of $3 \times 10^5 L_{\odot}$ from
\citet{choi08a} and an effective temperature of 3650 K from
\citet{massey06}, the location of \vycma\ on an HR diagram is
consistent with an age of of 8.2 Myr, based on an evolutionary track of
a 25 $M_{\odot}$ star of solar metallicity with an initial rotational
velocity of 300 \kms\ at its equator \citep{meynet03}.  This age
discrepancy between the motion time and age of \vycma\ may imply that
the relative proper motion between the molecular cloud and \vycma\ is
much smaller than its absolute proper motion or that the stellar
evolution model is not appropriate for \vycma. 

\begin{table}[H]
  \footnotesize
  \caption[]{Absolute position and proper motion of \vycma\ from different telescopes}
  \begin{center}
  \begin{tabular}{cccccc}
\hline \hline
Telescope & Epoch  &  R.A. (J2000)  & Dec. (J2000)                & \mux      &  \muy    \\   
          &  & (h~~~m~~~s)    & (\degr~~~\arcmin~~~\arcsec) & (\masy)   &  (\masy) \\
\hline
%--- mux
%--- 1997
Hipparcos$^a$ & 1991.25  &  07 22 58.3251 $\pm$ 0.0001 & $-$25 46 03.180 $\pm$ 0.003  & $+$8.86 $\pm$ 1.34 & $+$0.75 $\pm$ 3.25  \\
%--- 2007
Hipparcos$^b$ & 1991.25  &  07 22 58.3251 $\pm$ 0.0002 & $-$25 46 03.176 $\pm$ 0.003  & $+$5.72 $\pm$ 2.01 & $-$6.75 $\pm$ 4.47 \\
%--- 2008 PASJ (It said mu_ra, but I think it is mu_x)
VERA$^c$      & 2006.82  &  07 22 58.3264 $\pm$ 0.000? & $-$25 46 03.066 $\pm$ 0.00?  & $-$2.09 $\pm$ 0.16 & $+$1.02 $\pm$ 0.61  \\ 
%--- 2008 EVN IX meeting proceeding (position),  2008 PASJ (proper motion)
VLBA$^d$      & 2006.53  &  07 22 58.3259 $\pm$ 0.0007 & $-$25 46 03.070 $\pm$ 0.010  & $-$2.21 $\pm$ 0.06 & $+$2.29 $\pm$ 0.30  \\
\hline
  \end{tabular}
  \end{center}
\small{
a~ \citet{perryman97}. \\
b~ \citet{vanleeuwen07}. \\
c~ \citet{choi08a, choi08b}, where the position uncertainty is unknown. \\
d~ This paper.
} 
\label{tab:src_abspos}
\end{table}

%--- Fig  Geometry of \vycma's environment. 

\section{CONCLUSIONS}

We have measured the trigonometric parallax and proper motion of \vycma\
from VLBA observations of the variations of relative positions between
43 GHz SiO masers and the background source \Jtwofi.  The parallax of
0.83 $\pm$ 0.08 mas provides a distance of $1.20^{+0.13}_{-0.10}$ kpc,
which is consistent with that measured with 22 GHz \hho\ maser using the
VERA by \citet{choi08a}.  There can be little doubt that \vycma\ is
nearer than the value of 1.5 kpc generally used in the literature.

Using the VLA detection of the radio photosphere, using the SiO masers
as a phase reference, we determined the position of the central star
relative to the SiO masers.  This position is consistent with that
estimated from the distribution of maser spots with an uncertainty
better than 10 mas.  We found there are several spoke-like maser
features; most of these could be modeled by ballistic orbits, suggesting
that most spoke-like maser features are on the far side of the star, are
decelerating, pointing back to the central star.

The kinematics of SiO maser spots show a slow expansion.  After removing
the effects of expansion, we derived an absolute proper motion of the
central star.  A large discrepancy of tens of mas between the positions
of \vycma\ determined by optical and radio measurements is found, and we
suggest that optical position determinations are affected by the
scattering from circumstellar dust and cannot be used to locate the star
or its motion.

%--- Acknowledgement
\vskip 1 cm

We wish to thank the anonymous referee for very detailed and helpful
comments.  B. Zhang is supported by the National Science Foundation of
China under grant 10703010, 11073046, and the Knowledge Innovation
Program of the Chinese Academy of Sciences.

%\bibliographystyle{apj}
%\bibliography{ref}

\end{CJK*}
\end{document}